\documentclass[english]{article}
\usepackage{lmodern}

\usepackage[T1]{fontenc}
\usepackage[latin9]{inputenc}
\usepackage[a4paper]{geometry}
\usepackage{color}
\usepackage{babel}
\usepackage{natbib}

\usepackage{amsthm,amsmath}
\usepackage{mathrsfs}
\usepackage{bbm}
\usepackage{bm}
\usepackage{amssymb}
\usepackage{extarrows}
\usepackage{framed}
\usepackage{graphicx}
\usepackage{graphics}
\usepackage{comment}
\usepackage[section]{algorithm}
\usepackage{float}
\usepackage{multirow}
\usepackage{enumerate}
\usepackage{subcaption}
\usepackage[noend]{algpseudocode}
\usepackage{dsfont}
\usepackage[unicode=true,bookmarks=true,bookmarksnumbered=false,bookmarksopen=false,breaklinks=true,colorlinks=true,backref]{hyperref}
\usepackage{authblk}
\usepackage[dvipsnames,svgnames,x11names,hyperref]{xcolor}
\hypersetup{urlcolor=Peru,linkcolor=RoyalBlue,citecolor=RoyalBlue}
\makeatletter
\usepackage[normalem]{ulem}
\makeatother

\numberwithin{equation}{section}
\theoremstyle{plain}
\theoremstyle{remark}

\newtheorem{assump}{Assumption}[section]

\newtheorem{example}{Example}[section]

\title{An invitation to sequential Monte Carlo samplers}
\author[1]{Chenguang Dai}
\author[2]{Jeremy Heng}
\author[2]{Pierre E. Jacob \thanks{Corresponding author: pierre.jacob@essec.edu}}
\author[3]{Nick Whiteley}
\affil[1]{Citadel Securities}
\affil[2]{ESSEC Business School, Singapore}
\affil[3]{School of Mathematics, University of Bristol, UK}
\usepackage{setspace}
\date{}

\begin{document}
\maketitle

\begin{abstract}
  Statisticians often use Monte Carlo methods to approximate probability
  distributions, primarily with Markov chain Monte Carlo and importance
  sampling.  Sequential Monte Carlo samplers are a class of algorithms that
  combine both techniques to approximate distributions of interest and their
  normalizing constants. These samplers originate from particle filtering for
  state space models and have become general and scalable
  sampling techniques.  This article describes
  sequential Monte Carlo samplers and their possible implementations, 
  arguing that they remain under-used in statistics, despite their ability
  to perform sequential inference and to leverage parallel processing resources
  among other potential benefits.
\end{abstract}

\section{Introduction}\label{sec:introduction}

\paragraph{Motivation.} The numerical approximation
of probability distributions is ubiquitous in statistics, whether it concerns
the distribution of a test statistic, or conditional distributions of
parameters and latent variables given observed variables.  Monte Carlo methods,
originally developed in physics to compute specific expectations of interest,
have become fundamental in data analysis.  The complexity of distributions
encountered in the practice of statistics has increased, and the study of Monte
Carlo methods to approximate them has become more formal. Meanwhile, the
context surrounding the development of numerical methods has changed, with
computation becoming more parallel, users becoming accustomed to stochastic as
opposed to deterministic calculations, increasing interest in quantifying 
Monte Carlo errors, and a wider availability of modifiable
software packages. 

In this fast-changing landscape, it can be difficult for the
non-specialist to keep track of important developments in Monte Carlo
methods. This article shines some light on 
\emph{Sequential Monte Carlo samplers}, imported from
the signal processing literature to statistics in the early 2000s
\citep{chopin2002sequential,del2006sequential}.  These algorithms provide a
generic approach to sample from probability distributions,
can scale well with the dimension of the state space, provide
estimates of the associated normalizing constants, are well-adapted to 
sequential settings
and are largely amenable to parallel computing. 

We first recall Markov chain Monte Carlo
\citep[MCMC,][]{robert2011short,dunson2020hastings} and importance sampling 
 \citep[IS, Chapter
 9,][]{owen_2013} techniques and highlight some of their limitations.
Throughout the article we consider the task of sampling from a target distribution $\pi(dx) =
\gamma(x)dx/Z$ defined on a measurable space $(\mathsf{X}, \mathscr{X})$, with
unnormalized density $x\mapsto \gamma(x)$ and unknown
normalizing constant $Z=\int_{\mathsf{X}}\gamma(x)dx$, that we might also want to approximate. 

\paragraph{MCMC.} Given a $\pi$-invariant Markov transition
kernel $M$, an MCMC method generates a Markov chain $(x_t)_{t\geq 0}$ by sampling
$x_0$ from an initial distribution $\pi_0$ and iteratively sampling the next
state $x_t$ given $x_{t-1}$ from $M(x_{t-1},\cdot)$.  After discarding an initial
portion of the trajectory as ``burn-in'', the subsequent $T$ states form an
empirical approximation $T^{-1}\sum_{t=1}^T \delta_{x_t}(\cdot)$ of the target
distribution $\pi$ converging as $T\to\infty$ \citep{NummelinMCMC}.  Although
immensely successful \citep{diaconis2009markov}, MCMC methods suffer from
serious limitations: their iterative nature prevents straightforward
parallelization; tuning the Markov transition kernels to achieve a satisfactory
efficiency might be time-consuming; and the estimation of the normalizing constant $Z$
from MCMC runs alone is difficult.

Motivated by these considerations, a number of more elaborate algorithms,
or ``meta-algorithms'', emerged in the
1990s, such as parallel tempering \citep{geyer1991markov} where intermediate
distributions of varying complexity are introduced between $\pi_0$ and $\pi$, and
Markov chains targeting these distributions regularly exchange their states.
The introduction of a path of distributions also appeared in techniques
specifically designed to estimate $Z$, known as bridge sampling
and path sampling \citep{chen1997monte,gelman1998simulating}.

\paragraph{Importance sampling.} IS is a method to approximate
$\pi$ and $Z$ using samples from another distribution $\pi_0$.  Denoting $N$
independent samples from $\pi_0$ by $(x^n)_{n\in[N]}$, IS assigns weights
computed as $w^n=\gamma(x^n)/\pi_0(x^n)$ for each $n\in[N]=\{1,\ldots,N\}$.
This provides the estimator $Z^N = N^{-1} \sum_{n=1}^N w^n$ and the weighted
empirical measure $(N Z^N)^{-1} \sum_{n=1}^N w^n\delta_{x^n}(\cdot)$ that 
converge to $Z$ and $\pi$ as $N\to \infty$ under assumptions on the
ratio of densities $\pi(x)/\pi_0(x)$ \citep{owen_2013}.  The method is amenable to parallel
computation but its naive implementation can fail spectacularly when $\pi_0$ and $\pi$ are far apart
\citep{agapiou_2014,chatterjee2015sample}. This motivated
the introduction of ``bridging'' distributions between $\pi_0$ and $\pi$ 
and the combination of these intermediate distributions with MCMC moves 
was proposed in Annealed Importance Sampling \citep{neal2001annealed},
see also \citet{jarzynski_1997}.
These works are predecessors to sequential Monte Carlo samplers.

\paragraph{Particle filters.} State space models are a flexible
way of analyzing time series, where each observation 
is associated with a latent variable, and these latent variables form a
Markov chain.  In that context, importance sampling forms the basis of particle
filters \citep{gordon1993novel,kong1994sequential}, developed for
sequential state inference where the target $\pi_t$ at step $t$ is the
distribution of the $t$-th latent variable given observations available up
to time $t$.  Particle filters employ Monte Carlo principles 
that generalize
Kalman filters to non-linear, non-Gaussian state space models.  Many
variants have been put forward to deal with state space models of various
complexities \citep{capp2007review,chopin2020introduction}, 
for example with the addition of MCMC moves at each step 
\citep{gilks2001following}. 

\paragraph{SMCS.} 
By the late 1990s, various Monte Carlo algorithms incorporated ideas
from both importance sampling and MCMC, and the introduction of intermediate distributions
was already familiar. There were concurrent efforts to generalize 
particle filters to a much wider class of problems arising in statistics,
under the name of Iterated Batch Importance Sampling \citep{chopin2002sequential}
and SMC samplers \citep[SMCS,][]{del2006sequential}.
To avoid confusion with particle filters, we refer to SMC samplers as SMCS 
and we view them as
a generic alternative to MCMC and IS
for the approximation of $\pi$ and $Z$.
SMCS share the general structure of particle filters and we
elaborate further on the links in the appendices. 

\begin{figure}[th]
  \centering
  \includegraphics[width=1.0\textwidth]{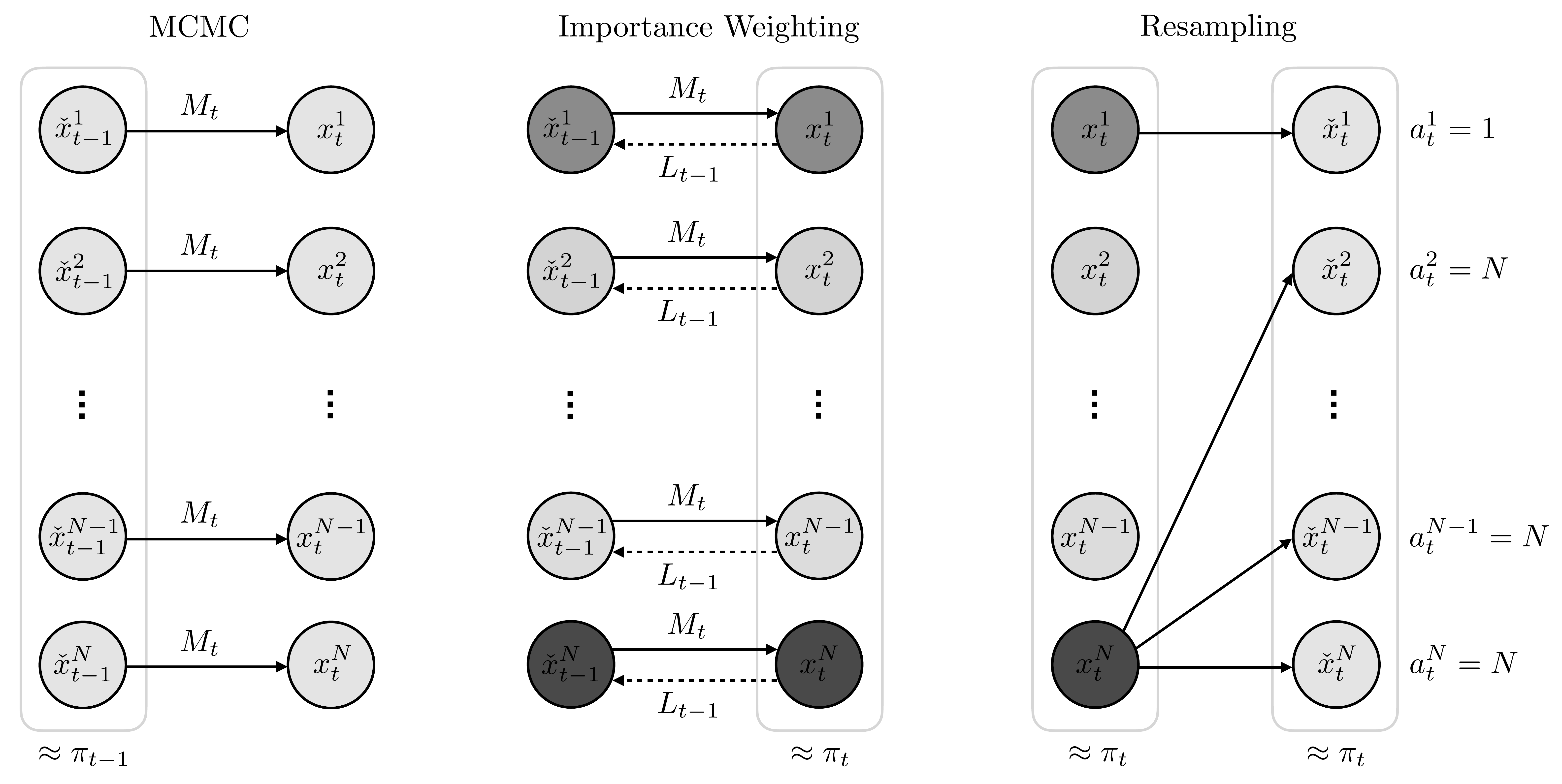}
  \caption{The three steps of SMCS. Particles $(\check{x}_{t-1}^n)_{n\in[N]}$ approximating $\pi_{t-1}$ evolve 
  separately according to an MCMC kernel $M_t$ (\emph{left}). To approximate
$\pi_t$, each pair $(\check{x}_{t-1}^n,x_t^n)$ is assigned an importance weight
$w_t(\check{x}_{t-1}^n,x_t^n)$, which depends on a backward kernel $L_{t-1}$
(\emph{middle}). The size of weights is depicted by the shade of grey. 
Equally weighted particles are then obtained by sampling from
the particles according to their weights (\emph{right}).}
  \label{fig:schematic}
\end{figure}

SMCS are
designed to approximate an arbitrary sequence of target distributions of fixed
dimension, that recovers the desired distribution $\pi$ as its last element,
and estimates of $Z$ are obtained as by-products. SMCS involve sequences of
bridging distributions or ``paths'', as in path sampling and parallel
tempering.  Specifically, SMCS generate a system of $N$ samples, termed
particles, that evolve through MCMC moves and importance weights, and interact
through resampling steps.  A schematic description of SMCS is provided in
Figure \ref{fig:schematic}.
The benefits of SMCS include the estimation of $Z$,
the ability to adaptively tune MCMC kernels, a large amenability to parallel
computing, and improved accuracy relative to plain or annealed IS.

\paragraph{Objectives and outline.} We aim to provide an accessible guide 
with useful references on SMCS for statisticians.
We do not assume familiarity with particle filters.  Our
presentation is self-contained, complementing Chapter 17 of
\citet{chopin2020introduction} with updated pointers to some
of the most significant advances.  Section \ref{sec:practice} describes 
SMCS and possible instantiations.  As SMCS involve IS steps, and since
the performance of one step of IS tends to deteriorate with the dimension, we
provide in Section \ref{sec:whybridginghelp} a clarification of the
role of bridging distributions.  In Section \ref{sec:theory} we present
methodological consequences of the fact that SMCS generate interacting particle
systems \citep{del2004feynman,del2013mean}, and not trajectories of Markov
chains as in MCMC.  We underline how parallel computers can be employed and how
the approximation error can be quantified.  In Section \ref{sec:objects}
we illustrate how SMCS translate into algorithms of practical importance in
statistics using simple examples, and Section \ref{sec:discussion} concludes.

\section{What are sequential Monte Carlo samplers?} \label{sec:practice}

We follow \citet{del2006sequential},
and describe the required specification of paths,
of forward and backward Markov transition kernels and of various algorithmic components. 
As we will see, many of these choices can be implicitly or adaptively made.

\subsection{Algorithmic description}\label{sec:generic_smc}

A generic, non-adaptive sequential Monte Carlo sampler is described in Algorithm \ref{alg:smc}. 
As input, a sequence of $T+1$ distributions  
$\pi_0,\ldots,\pi_T$ is introduced, with each 
$\pi_t(dx)=\gamma_t(x)dx/Z_t$ defined on the same space $(\mathsf{X}, \mathscr{X})$,  
where $\gamma_t(x)$ denotes an unnormalized density, 
and $Z_t=\int_{\mathsf{X}}\gamma_t(x)dx$ a normalizing constant (with $Z_0=1$). 
We assume that we can sample from $\pi_0$ to initialize the algorithm (see Step 1(a)) and that the terminal distribution $\pi_T$ is precisely the target distribution $\pi$. 

Next we introduce two sequences of Markov kernels. 
In the sequence $(M_t)_{t\in[T]}$ of ``forward'' kernels, each $M_t$ is designed to leave $\pi_t$ invariant, at
least approximately.
The forward kernel $M_t$ is used to sample $x_t$ given $x_{t-1}$
at the $t$-th step of the algorithm (see Step 2(b) and left of Figure \ref{fig:schematic}). 
At this stage, one could imagine an importance sampling step
with the proposed $x_t$ to approximate $\pi_t$. 
However, for most choices of $M_t$, the marginal density of the proposed state 
$q_t(x_t) = \int_{\mathsf{X}} \pi_{t-1}(dx_{t-1})M_t(x_{t-1},x_t)$
would be intractable, and importance weights could not be computed. 
To circumvent this issue, the key idea  
is to define an importance sampling step on the space of $(x_{t-1},x_t)$,
by introducing the sequence $(L_{t-1})_{t\in[T]}$ of ``backward'' kernels \citep{neal2001annealed} (see Step 2(c) and middle of Figure \ref{fig:schematic}). 
Defining the proposal $\pi_{t-1}(dx_{t-1})M_t(x_{t-1},dx_t)$ and the target $\pi_t(dx_t) L_{t-1}(x_t,dx_{t-1})$ on the joint space, the weight function 
\begin{equation}
	w_t(x_{t - 1}, x_t) = \frac{\gamma_t(x_t)L_{t - 1}(x_t, x_{t - 1})}{\gamma_{t - 1}(x_{t - 1})M_t(x_{t - 1}, x_t)}
	\label{eq:smc-incremental-weight}
\end{equation}
can be made tractable even if the marginal distribution of $x_t$ 
is intractable and even if $M_t$ has an intractable transition density,
by choosing $L_{t-1}$ appropriately. Indeed the backward kernel $L_{t-1}$
is introduced 
specifically so that the importance weight in \eqref{eq:smc-incremental-weight} achieves 
a practical compromise between ease of numerical evaluation and variance.
The joint target $\pi_t(dx_t)L_{t-1}(x_t,dx_{t-1})$ admits $\pi_t$ as marginal on $x_t$, for any choice of backward kernel $L_{t-1}$, therefore \eqref{eq:smc-incremental-weight} provides a valid importance weight. 
As described in \citet[Section 3.3,][]{del2006sequential}, given $\pi_{t-1}$, $\pi_t$, and $M_t$,
one should ideally select $L_{t-1}$ to keep the variance of $w_t(x_{t-1},x_t)$ under the joint proposal $\pi_{t-1}(dx_{t-1})M_t(x_{t-1},dx_t)$ small. 
Minimal variance is attained by $L_{t-1}(x_t,x_{t-1})=\pi_{t - 1}(x_{t -
1})M_t(x_{t - 1}, x_t)/q_t(x_t)$, the backward transition under the joint
proposal, that reduces to having $w_t(x_{t-1},x_t) = \gamma_t(x_t)/q_t(x_t)$,
which, again, would be typically intractable.
In other words tractability of importance sampling through 
operating on the joint space of $(x_{t-1},x_{t})$ comes with an increase of variance.

Overall, the algorithm propagates $N$ particles $x_t^{1:N}=(x_t^1,\ldots,x_t^N)$
using the forward kernels $(M_t)$, and assigns to the particles some weights that depend
on $(\pi_t)$, $(M_t)$ and $(L_{t-1})$. 
The above reasoning could be carried out recursively and the final weights would
be obtained as the product over time of the weights computed at each time step \citep{neal2001annealed}.
One might then worry about the variance of the weights after $T$ steps.
The resampling step (see Step 2(a) and right of Figure \ref{fig:schematic}) helps to mitigate this issue.
According to the magnitude of their weights, 
some particles are discarded and others duplicated, maintaining a fixed population size of $N$.
Resampling involves a distribution $r(\cdot|w^{1:N})$ of ancestor indices $a^{1:N}$ in $[N]^N$, parametrized by 
a vector $w^{1:N}$ of probabilities. 
The simplest
resampling scheme is called multinomial resampling \citep{gordon1993novel},
where $a^{1:N}$ are independent Categorical variables on $[N]$ with 
probabilities $w^{1:N}$. At step $t$ of the algorithm, 
$N$ particles are obtained by propagating 
particles with indices $(a_{t-1}^{n})_{n\in[N]}$
generated from $r(\cdot|w_{t-1}^{1:N})$.
\citet{gerber_2017} provide recent discussions on resampling schemes, 
\citet{whiteley2016role} study the stabilizing effect of resampling, and  
\citet{gerber2015sequential} consider the use of quasi-random numbers. 
One could resample a different number of particles $N_t$ at each step $t$,
and the optimal allocation of these numbers is discussed in \citet{lee2015variance}.
Resampling is the key difference between Annealed Importance Sampling \citep{neal2001annealed}
and SMCS, and we recover the former by omitting the resampling steps, i.e. $\check{x}_{t-1}^n = x_{t-1}^n$ in Step 2(a).
Resampling makes the SMCS output non-differentiable with respect to the input,
which has motivated works such as \citet{corenflos2021differentiable}.

The output of SMCS includes weighted particles $(w_t^n, x_t^n)_{n\in[N]}$ 
approximating $\pi_t$, 
in the sense that $\pi_t^N(\varphi)=\sum_{n\in[N]} w_t^n \varphi(x_t^n)$ 
converges to $\pi_t(\varphi)=\int_{\mathsf{X}} \varphi(x_t)\pi_t(dx_t)$, for suitable $\varphi:\mathsf{X}\rightarrow\mathbb{R}$, 
as $N\rightarrow\infty$.  
Another output of the algorithm is an unbiased and consistent normalizing constant estimator $Z^N_t$, 
computed using the unnormalized weights.

\begin{algorithm}
	\caption{Sequential Monte Carlo sampler}\label{alg:smc}
\begin{flushleft}
	\textbf{Input:} sequence of distributions $(\pi_t)$, forward Markov kernels $(M_t)$, backward Markov kernels $(L_t)$,
	resampling distribution $r(\cdot|w^{1:N})$ on $[N]^N$ where $w^{1:N}$ is an $N$-vector of probabilities.
\begin{enumerate}
\item Initialization.
\begin{enumerate}
	\item Sample particle $x_0^n$ from $\pi_0(\cdot)$ for $n\in[N]$ independently.
	\item Set $w_0^n = N^{-1}$ for $n\in[N]$.
\end{enumerate}
\item For $t \in [T]$, iterate the following steps.
\begin{enumerate}
	\item Sample ancestor indices $(a_{t-1}^{n})_{n\in[N]}$ from $r(\cdot|w^{1:N}_{t-1})$,
	 \item[] and define $\check{x}_{t-1}^n = x_{t-1}^{a_{t-1}^n}$ for $n\in[N]$.
    \item Sample particle $x_t^n\sim M_t(\check{x}_{t - 1}^n,\cdot)$ for $n\in[N]$.
	\item Compute weights $w_t(\check{x}_{t - 1}^n, x_t^n)=\frac{\gamma_t(x_t^n)L_{t - 1}(x_t^n, \check{x}_{t - 1}^n)}{\gamma_{t - 1}(\check{x}_{t - 1}^n)M_t(\check{x}_{t - 1}^n, x_t^n)}$ for $n\in[N]$,
\item[] and set $w_t^n \propto w_t(\check{x}_{t - 1}^n, x_t^n)$ such that $\sum_{n\in[N]} w_t^n = 1$.
\end{enumerate}
\end{enumerate}
\textbf{Output:} weighted particles $(w_t^n,x_{t}^n)_{n\in[N]}$ approximating $\pi_t$,
and estimator $Z^N_t = \prod_{s=1}^t N^{-1} \sum_{n\in[N]} w_s(\check{x}_{s-1}^n,x_s^n)$ of $Z_t$ for $t\in[T]$.
\end{flushleft}
\end{algorithm}

\subsection{Paths of distributions}
\label{subsec:choosepath}

The initial distribution $\pi_0(dx) =
\gamma_0(x)dx/Z_0$ and target distribution $\pi(dx) = \gamma(x)dx/Z$ are
considered inputs of the problem.  We consider the choice of a path 
$\pi_t(dx)=\gamma_t(x)dx/Z_t$ for $t\in[T]$, where the number of
distributions $T$ can be user-specified or determined adaptively (see 
Section \ref{subsec:choosenexttemp}).  

\paragraph{Geometric path.} 
A popular choice is the geometric path 
\begin{equation}
	\gamma_t(x) = \gamma_0(x)^{1-\lambda_t}\gamma(x)^{\lambda_t}, \label{eq:geometricpath}
\end{equation}
defined by a sequence $0= \lambda_0<\lambda_1<\dots<\lambda_T=1$, 
often referred to as inverse temperatures  \citep{kirkpatrick1983optimization}.  
The unnormalized density $\gamma_t(x)$ and its gradient can be evaluated
pointwise if it is possible to evaluate $\gamma_0(x)$ and $\gamma(x)$
and their gradients.
If $\pi_0$ is a proper prior and $\pi$ the posterior
distribution, then the geometric path corresponds to raising the likelihood
to a power.  There could be statistical reasons to consider the
resulting ``tempered'' posteriors, for example see
\citet{HolmesWalker}
and references therein.
A run of SMCS using \eqref{eq:geometricpath} provides
approximations of tempered posteriors over a desired sequence of exponents.
Geometric paths can be generalized into q-paths,
where the geometric mean in \eqref{eq:geometricpath}
is replaced by a power mean \citep{whitfield2002generalized};
see \citet{syed2021parallel} in the context of parallel tempering.

\paragraph{Path of partial posteriors.} 
In the Bayesian setting, with a prior distribution $\pi_0(dx)=p(dx)$
and the target distribution $\pi(dx) = p(dx|y_{1:T})$ defined as the posterior
based on observations $y_{1:T}=(y_1,\ldots,y_T)$,
\citet{chopin2002sequential} proposed SMCS, then termed Iterated Batch Importance Sampling,
applied to the sequence of ``partial posteriors''
$\pi_t(dx)=p(dx|y_{1:t})$ for $t\in[T]$.  
The procedure provides a richer analysis compared to the
approximation of $p(dx|y_{1:T})$ alone, helping one  
to visualize how the posterior distribution and the marginal likelihood 
evolve as data points are assimilated.
Concepts such as Bayesian updating, sequential analysis and coherency are 
often presented as central in Bayesian theory (e.g. Section 2.4.4 of
\citet{bernardo2009bayesian} or \citet{hooten2019}),
and yet it is common in applied Bayesian analysis
to examine only the posterior distribution given all the data.
SMCS on the path of partial
posteriors enable the consideration of the dynamics
of Bayesian inference, as will be illustrated in Section \ref{sec:objects}.  
This path is also key to the
assessment of sequential predictive performance,
such as with the Hyv\"arinen score \citep{dawid2015bayesian,shao2019bayesian},
and plays a special role in Bayesian sequential experimental design
\citep{drovandi2014sequential,cuturi2020noisy}. 
The path of partial posteriors is particularly appealing for time series,
and when combined with 
particle filters for non-linear state space models,
the technique is known as SMC$^2$ \citep{chopin2013smc2,fulop2013efficient}.

Improved performance can be obtained by introducing a geometric path between
successive partial posteriors. In the presence of improper priors, the
initial distribution cannot be set as the prior. As an alternative, 
one can start the algorithm using 
a geometric path between some proper distribution and the
posterior distribution given enough observations for it to be proper.

\paragraph{Path of truncated distributions.} 
In rare event estimation, the task 
is to approximate the probability of a set $A\in\mathscr{X}$ under a distribution $\mu(dx)=\mu(x)dx$ 
defined on $(\mathsf{X}, \mathscr{X})$. 
Following \citet{cerou2012sequential}, we consider sets 
$A=\{x \in \mathsf{X}: \Phi(x)\geq \ell \}$ for
a function $\Phi:\mathsf{X}\to \mathbb{R}$ and a level $\ell\in\mathbb{R}$. 
We can define 
\begin{equation}\label{eqn:truncated_distributions}
	\gamma_t(x) = \mu(x)\mathbb{I}_{A(\ell_t)}(x),
\end{equation}
where $-\infty = \ell_0 < \ell_1 < \ldots < \ell_T = \ell$ is a sequence of levels, and 
$\mathbb{I}_{A(l)}(x)$ denotes the indicator function on the set 
$A(l) = \{x \in\mathsf{X}: \Phi(x)\geq l\}$. 
This defines a path of distributions that gradually truncates $\pi_0(dx)=\mu(dx)$ to 
$\pi(dx)=\mu(dx)\mathbb{I}_A(x)/Z$, whose normalizing constant 
$Z=\mu(A)$ is the probability of interest. 

In the Bayesian setup where $\pi_0$ and $\pi$ denote a (proper) prior and posterior, respectively,  
nested sampling \citep{skilling2006nested} 
represents the marginal likelihood as $Z=\int_0^\infty \pi_0(A(l))dl$ with $A(l)$ defined by  
level sets of the likelihood function $\Phi(x)=\gamma(x)/\gamma_0(x)$. 
This identity is leveraged by \citet{salomone2018unbiased} to apply SMCS 
with the path \eqref{eqn:truncated_distributions}.
Approximate Bayesian computation 
provides another setting where a sequence of truncated distributions,
indexed by a ``tolerance'' parameter, 
can be estimated by SMCS 
\citep{sisson2007sequential,del2012adaptive}.

\paragraph{Path of least coding effort.} In anticipation of the choice of
forward Markov kernels $(M_t)$, we might want to introduce a path of
distributions such that the associated Markov kernels are readily available.
One might already have an MCMC algorithm that targets $\pi$, for example a
Gibbs sampler that exploits specific aspects of $\pi$.  To reduce
implementation effort, we can then introduce a path $(\pi_t)$ designed so that
only slight modifications to that MCMC algorithm are required. 
For example the implementation of Langevin or Hamiltonian Monte Carlo to target 
any distribution on the geometric path 
requires minimal modifications relative to the original target $\pi$.

Another example can be found in \citet{rischard2018unbiased},
in the context of logistic regression. Assuming 
a Normal prior on the regression coefficients,
the P{\'o}lya--Gamma Gibbs (PGG) sampler of \citet{polson2013bayesian}
can be employed to target the posterior distribution,
for any matrix of covariates $x$ and binary outcome vector $y$.
We can introduce a path of 
posterior distributions $\pi_t$ corresponding to the use of scaled covariates $\lambda_t x$
instead of $x$, with $\lambda_t \in[0,1]$.
The appeal is that the same implementation 
of PGG, given inputs $\lambda_t x$ and $y$, provides a forward kernel $M_t$ for each distribution $\pi_t$.
A similar approach was considered for probit regressions
in \citet{smcbayesiancomputation}.

There is much freedom in the choice of paths,
so that various settings and goals can be accommodated.
Sequences of distributions can be further generalized to sets of distributions indexed
by trees, with applications to Bayesian hierarchical models in \citet{lindstenDaCSMC}.  

\subsection{Forward and backward Markov kernels}
\label{subsec:choosemarkov}
Given a path $(\pi_t)$, the SMCS user must select forward and backward kernels, 
$(M_t)$ and $(L_{t})$. 
In view of Algorithm \ref{alg:smc}, one must be able to sample from $M_t(x_{t-1},\cdot)$ 
and to evaluate $w_t(x_{t-1},x_t)$ 
in \eqref{eq:smc-incremental-weight}.
We would set $M_t(x_{t-1},dx_t)=\pi_t(dx_t)$ if perfect samples from $\pi_t$ could be obtained, and 
by defining $L_{t-1}(x_t,dx_{t-1})=\pi_{t-1}(dx_{t-1})$ the weight 
would simplify to $Z_t/Z_{t-1}$, leading to an estimator of $Z_t$ with zero variance. 
This section presents more practical choices.

\paragraph{Exact MCMC moves.}
We can exploit the vast literature on
MCMC to design $M_t$ as a $\pi_t$-invariant kernel. 
Although such choices typically do not admit tractable transition densities, 
the weight in \eqref{eq:smc-incremental-weight} can be tractable if the backward 
kernel $L_{t-1}$ is chosen judiciously. 
Following \citet{jarzynski_1997,neal2001annealed}, 
$L_{t-1}$ can be selected as the time reversal of $M_t$, 
i.e. $L_{t-1}(x_t,x_{t-1})=\pi_t(x_{t-1})M_t(x_{t-1},x_t)/\pi_t(x_t)$, 
leading to  
the weight $\gamma_t(x_{t-1})/\gamma_{t-1}(x_{t-1})$. 
When the distributions $\pi_{t-1}$ and $\pi_t$ are close, the time reversal provides an approximation of the backward transition 
$L_{t-1}(x_t,x_{t-1})=\pi_{t - 1}(x_{t -1})M_t(x_{t - 1}, x_t)/q_t(x_t)$ yielding minimal variance. 
\citet[Section 3.3]{del2006sequential} provide 
more discussions on the choice of $(L_{t})$ given $(M_t)$. 

SMCS can also accommodate 
kernels $M_t$
that are not $\pi_t$-invariant, 
while preserving consistency of SMC estimates. 
We provide an example that shows how to remove time-discretization biases without 
resorting to Metropolis--Rosenbluth--Teller--Hastings corrections. 
 
\paragraph{Unadjusted Langevin moves.}
For problems on $\mathsf{X}=\mathbb{R}^d$, we can select forward kernels based on the unadjusted Langevin algorithm \citep[ULA,][]{grenander1994representations}:
\begin{align}
  M_t(x_{t-1},dx_t) = \mathcal{N}\left(x_t;x_{t-1} + \frac{\varepsilon}{2} \Omega \nabla \log \pi_t (x_{t-1}), 
\varepsilon\Omega\right)dx_t,
\end{align}
where $z\mapsto\mathcal{N}(z;\mu,\Sigma)$ denotes the density of a Normal distribution 
with mean vector $\mu$ and covariance matrix $\Sigma$,
$\varepsilon>0$ denotes a step size, and $\Omega\in\mathbb{R}^{d\times d}$ is a positive definite preconditioning matrix.
In general the ULA transition does not leave $\pi_t$ invariant for any $\varepsilon>0$. 
When an acceptance correction step is added to enforce $\pi_t$-invariance, 
the resulting method is known as the Metropolis-adjusted Langevin algorithm (MALA).
In SMCS one can account for the time-discretization using importance sampling
instead. 
The reversibility of the underlying continuous-time Langevin diffusion suggests the choice 
$L_{t-1}(x_t, dx_{t-1}) = M_t(x_{t},dx_{t-1})$ for sufficiently 
small $\varepsilon$ \citep{nilmeier2011nonequilibrium}. 
With these choices, the weight \eqref{eq:smc-incremental-weight} is tractable,
and approaches $\gamma_t(x_{t-1})/\gamma_{t-1}(x_{t-1})$ as $\varepsilon\to 0$.
The tractability of ULA kernels
as an alternative to MALA kernels within SMCS was exploited in the controlled sequential Monte Carlo approach \citep{heng2020controlled}, 
which optimizes over the path of distributions $(\pi_t)$ and forward kernels $(M_t)$
to improve performance, and in the Schr\"odinger bridge sampler \citep{bernton2019schr} that fixes $(\pi_t)$ 
and optimizes over $(M_t)$ and $(L_{t-1})$ for similar purposes. 
As an alternative, the appendices describe the
use of unadjusted Hamiltonian Monte Carlo (HMC) moves within SMCS. 
One can also design the forward kernel $M_t$ as a deterministic map that transports $\pi_{t-1}$ to $\pi_t$, and choose the backward kernel $L_{t-1}$ as the inverse map \citep{vaikuntanathan2008escorted,heng2021gibbs,everitt2020sequential,arbel2021annealed,matthews2022continual}.

\paragraph{Tuning parameters.} 
Having chosen Markov kernels, there might be some algorithmic parameters
to tune.  Firstly, it is often worthwhile to iterate the chosen Markov kernel 
more than once at each step of SMCS. 
For $\pi_t$-invariant forward kernels with time reversals as
backward kernels, iterating the forward kernel can be done without modification
of the weights.  For unadjusted kernels additional care might be required.
When we iterate MCMC moves, or when we perform moves that involve
intermediate steps such as HMC, it can be advantageous to exploit 
all intermediate samples \citep{wastefreeSMC}; see also the appendices.

Each Markov 
kernel may further depend on parameters such as step sizes, 
or preconditioning matrices.  
An appealing specificity of SMCS, relative to classical MCMC, is that
approximations of the previous and current bridging distributions are available
and can be used to inform the choice of parameters.
For example one can select $\Omega$ as the estimated covariance of bridging
distributions for random walk or MALA moves \citep{chopin2002sequential}.  
\citet{fearnhead2013adaptive} offer a generic recipe to automate such
tuning procedures. \citet{kostov2016,buchholz2018adaptive} consider specifically HMC kernels,
and \citet{schafer2013sequential,south2019sequential} discuss strategies to
adapt independent proposals on discrete and continuous spaces.
While these adaptation rules will not affect
consistency properties of SMCS as $N\to\infty$ \citep{beskos_etal_2016_aap},
they may not preserve the unbiasedness property of normalizing constant
estimators. When this unbiasedness matters,
for example in particle MCMC methods as in Section
\ref{subsec:smallsmc}, one can run an adaptive SMCS, record the obtained
tuning parameters and run a second, non-adaptive sampler, for an approximate
two-fold increase in computing cost.

\subsection{Progressing through a path of distributions } 
\label{subsec:choosenexttemp}

The user also needs to address the choice of the number of
distributions $T$ and of the particular distributions $\pi_t$ along a given path.  
In the case of a geometric path \eqref{eq:geometricpath}, 
one needs a specific choice of inverse temperatures
$(\lambda_t)_{t\in[T]}$.  We could simply pre-specify $T$
and select $\lambda_t=(t/T)^p$ for
$t\in[T]$ and some exponent $p>0$, informed by preliminary
runs. The following describes a common procedure that specifies $T$ and
$(\lambda_t)_{t\in[T]}$ adaptively. For clarity, we 
consider only the setting where the forward kernel
$M_t$ is $\pi_t$-invariant and the backward kernel $L_{t-1}$ 
is its time reversal, leading to the weight  
\begin{align}\label{eqn:AIS_weight_geo}
	w_t(x_{t-1}) = \frac{\gamma_t(x_{t-1})}{\gamma_{t-1}(x_{t-1})} 
	= \frac{\gamma(x_{t-1})}{\gamma_0(x_{t-1})}^{\lambda_t - \lambda_{t-1}}.
\end{align}
As particle weights do not depend on their states at step $t$ in this setting, one 
can perform weighting (Step 2(c)) and resampling (Step 2(a)) before applying Markov moves 
(Step 2(b)) to promote sample diversity in Algorithm \ref{alg:smc}. 
Equation \eqref{eqn:AIS_weight_geo} can be seen as an importance weight targeting $\pi_t$ using proposed samples 
from $\pi_{t-1}$. Suppose $\pi_{t-1}$ has been determined by 
some $\lambda_{t-1}\in[0,1)$ and we seek 
$\lambda_t\in(\lambda_{t-1},1]$ so that $\pi_t$ 
can be well-approximated by $\pi_{t-1}$ through importance sampling. 
We can control the performance by keeping the 
$\chi^2$-divergence small \citep{agapiou_2014}, where
\begin{align}\label{eqn:chi_squared}
	\chi^2(\pi_t|\pi_{t-1}) = \int_{\mathsf{X}}\left(\frac{\pi_t(x)}{\pi_{t-1}(x)}-1\right)^2\pi_{t-1}(dx) 
	= \frac{\int_{\mathsf{X}} w_t(x)^2\pi_{t-1}(dx)}{\left(\int_{\mathsf{X}} w_t(x)\pi_{t-1}(dx)\right)^2}-1.
\end{align}
Instead of fixing $\chi^2(\pi_t|\pi_{t-1})$ to a desired level, it is more convenient to 
work with $\varrho_t(\lambda_t)=(1+\chi^2(\pi_t|\pi_{t-1}))^{-1}$ as this quantity takes values in $[0,1]$. 
Given samples $(x_{t-1}^n)_{n\in[N]}$ approximating $\pi_{t-1}$, 
an approximation of $\varrho_t(\lambda_t)$ is given by 
$\hat{\varrho}_t(\lambda_t) = \textrm{ESS}_t(\lambda_t)/N$, where 
\begin{align}\label{eqn:ESS_lambda}
	\textrm{ESS}_t(\lambda_t) = \frac{\left(\sum_{n=1}^Nw_t(x_{t-1}^n)\right)^2}{\sum_{n=1}^Nw_t(x_{t-1}^n)^2} 
	= \frac{\left(\sum_{n=1}^N(\gamma/\gamma_0)(x_{t-1}^n)^{\lambda_t-\lambda_{t-1}}\right)^2}
	{\sum_{n=1}^N(\gamma/\gamma_0)(x_{t-1}^n)^{2(\lambda_t-\lambda_{t-1})}}.
\end{align}
This effective sample size (ESS) \citep{kong1994sequential}
takes values in $[1,N]$, achieving the lower bound when one sample 
holds all the weight and the upper bound when all samples have equal weights. 

If $\hat{\varrho}_t(1)$ is greater than a pre-specified threshold
$\kappa\in(0,1)$, we set $\lambda_t=1$. 
Otherwise, we can solve for $\lambda_t\in(\lambda_{t-1},1)$ such that
$\hat{\varrho}_t(\lambda_t)$ is equal to $\kappa$ 
\citep{jasra2011inference}. 
As this enforces the $\chi^2$-divergence between
successive distributions to be approximately $\delta = \kappa^{-1}-1$ 
in the large $N$ regime, higher thresholds lead to more bridging distributions $T$.
The search for $\lambda_t$ can be implemented using the bisection method on the
interval $[\lambda_{t-1},1]$ as the function $\hat{\varrho}_t(\lambda_t)$ is
strictly decreasing \citep[Lemma 3.1]{beskos_etal_2016_aap}.  The cost of this
procedure is negligible as evaluations of \eqref{eqn:ESS_lambda} are
inexpensive once $(\gamma/\gamma_0)(x_{t-1}^n)$ have been pre-computed. 
Note that when the path is fixed, adaptively
resampling whenever the ESS falls below a threshold does not alter unbiasedness
\citep{whiteley2016role}.

Modifications and alternatives to the ESS criterion are proposed in  
\citet{cornebise2008adaptive,zhou2016toward,huggins2019sequential},
and can be used to select intermediate distributions. An alternative
method to determine $(\lambda_t)$ adaptively is described in \citet{nguyen2015efficient}. 
In any case, starting with a fixed or an adaptive schedule,
we can re-run SMCS with additional intermediate steps to improve 
performance without necessarily increasing the number of particles, as
mentioned in Section \ref{subsec:smallsmc} and the appendices.

\section{Effect of bridging distributions} \label{sec:whybridginghelp}

\subsection{The curse of dimension for importance sampling}

The cost of IS estimators is
related to the discrepancy between the proposal and target distributions,
as measured by the  $\chi^2$ or KL divergence
\citep{agapiou_2014,chatterjee2015sample}. For example, the number of samples
needed for an estimator of $Z$ to achieve a given variance is proportional to that $\chi^2$-divergence.
As the dimension $d\in\mathbb{N}$ of $\mathsf{X}$ grows, 
the $\chi^2$ and KL divergences between
$\pi_0$ and $\pi$ would often increase exponentially with $d$. Since each step
of SMCS involves importance sampling, 
concerns about their performance in high dimension are understandable. 
Remarkably, SMCS can deliver
reliable estimates for problems in high
dimension; see e.g. the applications to inverse problems in
\citet{kantas2014sequential}, 
and to spatio-temporal models in 
\citet{naessethnestedSMC}.
We propose simple elements to explain this operational
success.  

\subsection{Variance of the normalizing constant estimator}\label{subsec:norm_const_var}

We focus on the geometric path
\eqref{eq:geometricpath}, forward $\pi_t$-invariant kernels $(M_t)$, and 
backward kernels $(L_{t})$ taken as their time reversals.
We consider the variance of the normalizing constant estimator
\begin{equation}
\label{eq:productestimator}
{Z}_{T}^N = \prod_{t = 1}^T\frac{1}{N}\sum_{n = 1}^Nw_t(X_{t-1}^n),
\end{equation} 
produced by the modification of Algorithm \ref{alg:smc} described in Section
\ref{subsec:choosenexttemp} to promote sample diversity.
\cite{cerou2011nonasymptotic} established a formula for this estimator's non-asymptotic
variance in the setting of Feynman--Kac formulae. Our first step is
to make a simplifying assumption that allows
us to capture some of the essence of 
\cite{cerou2011nonasymptotic} with only simple calculations. 
\begin{assump}\label{assump:perfectly_mixing}
For all $t\in[T]$, the forward kernel is perfect: $M_t(x_{t-1},dx_t)=\pi_t(dx_{t})$.
\end{assump}
Our priority here is exposition rather than 
realism, but in practice if $M_t$ is taken as multiple iterations of 
an ergodic kernel targeting $\pi_t$, 
then Assumption \ref{assump:perfectly_mixing} essentially holds if the number of iterates
is large enough.
Using the unbiased property
of $Z_T^N$ and the identity
$Z=\prod_{t=1}^TZ_t/Z_{t-1}=\prod_{t=1}^T\{\int_{\mathsf{X}}w_t(x_{t-1})\pi_{t-1}(dx_{t-1})\}$,
a calculation shows that 
\begin{equation}
\label{eq:productvariance}
\text{Var}\left[\frac{{Z}_{T}^N}{Z}\right] = \prod_{t = 1}^T\left[1 + \frac{\chi^2(\pi_t|\pi_{t-1})}{N}\right] - 1.
\end{equation}
Hence $\chi^2$-divergences between
consecutive distributions appear in the variance of $Z_T^N/Z$. 

\subsection{Scaling the number of bridging distributions with dimension}

Next we introduce a
sequence of sampling problems, indexed by $d\in\mathbb{N}$. 
We will specify the inverse
temperatures $(\lambda_t)$ in a way that possibly depends on $d$. 
The next assumption captures the idealized performance of the
adaptive procedure of Section \ref{subsec:choosenexttemp} as $N\to\infty$,
and serves to dispense with technical subtleties that would arise 
if the number of distributions $T$ was random. 

\begin{assump}\label{assump:chisq}
For all $t\in[T-1]$, $\pi_{t-1}$ and 
$\pi_t$ satisfy $\chi^2(\pi_t|\pi_{t-1})\leq \delta$ 
for some pre-specified $\delta>0$ which is independent of $d$, and such that $\chi^2(\pi|\pi_{0})>\delta$.
\end{assump}
 
\begin{assump}\label{assump:T}
There exists $\alpha>0$ such that $T=O(d^{\alpha})$ as $d\rightarrow\infty$.
\end{assump}

The above assumption postulates how the number of bridging distributions $T$
scales with dimension $d$. 
The fact that Assumptions \ref{assump:chisq}-\ref{assump:T} can hold simultaneously 
will be illustrated through examples.
Since the $\chi^2$-divergence between successive distributions is fixed as $\delta=\kappa^{-1}-1$
under Assumption \ref{assump:chisq}, the relative variance in \eqref{eq:productvariance} is equal to $(1+\delta/N)^T-1$. 
As $d\rightarrow\infty$ and hence $T\rightarrow\infty$, to ensure stability of the estimator 
\eqref{eq:productestimator}, we can choose the number of particles $N$ such that 
$N=O(T)$ to keep the relative variance of a constant order (since $\lim_{N\to\infty}(1+\delta/N)^{N}=\exp(\delta)$). 
Therefore the overall cost of this idealized SMCS measured in terms of density evaluations 
would be $O(T^2)=O(d^{2\alpha})$, that is polynomial in $d$, under Assumption \ref{assump:T}. 
We next consider Assumptions \ref{assump:chisq}-\ref{assump:T} on a Normal example. 

\begin{example}
Set $\pi_0(dx)=\mathcal{N}(x;\mu_0,\Sigma)dx$ and 
$\pi(dx)=\mathcal{N}(x;\mu,\Sigma)dx$ for some $\mu_0,\mu\in\mathbb{R}^d$ 
and $\Sigma\in\mathbb{R}^{d\times d}$. Each distribution along the geometric path \eqref{eq:geometricpath} 
is Normal $\pi_t(dx)=\mathcal{N}(x;\mu_t,\Sigma)dx$ with mean vector 
$\mu_t = \mu_0 + \lambda_t(\mu-\mu_0)$ for $t\in[T]$.  
The $\chi^2$-divergence between 
successive distributions can be computed in closed-form:
$\chi^2(\pi_t|\pi_{t-1})=\exp((\lambda_t-\lambda_{t-1})^2|\mu-\mu_0|^2_{\Sigma^{-1}})-1$,
where $|\mu-\mu_0|_{\Sigma^{-1}}=\sqrt{(\mu-\mu_0)^\top\Sigma^{-1}(\mu-\mu_0)}$. 
Under the specification
\begin{equation}\label{eqn:normal_numbridges}
T=\lceil |\mu-\mu_0|_{\Sigma^{-1}}/\sqrt{\log(1+\delta)}\rceil,
\quad 
\lambda_t = t\sqrt{\log(1+\delta)}/|\mu-\mu_0|_{\Sigma^{-1}} \text{ for } t\in[T-1], 
\end{equation}
where $\lceil\cdot\rceil$ denotes the ceiling function, 
Assumption \ref{assump:chisq} is satisfied.
Using the bound $|\mu-\mu_0|_{\Sigma^{-1}}\leq \Lambda_{\min}(\Sigma)^{-1/2}|\mu-\mu_0|$, 
where $\Lambda_{\min}(\Sigma)$ denotes the minimum eigenvalue of $\Sigma$, 
it follows from \eqref{eqn:normal_numbridges} that $T=O(\sqrt{d})$ if 
$\Lambda_{\min}(\Sigma)$ is uniformly bounded away from zero and $|\mu-\mu_0|$ is $O(\sqrt{d})$, both
as $d\rightarrow\infty$. In this situation, Assumption \ref{assump:T} holds with $\alpha=1/2$. 
\end{example}

To address less specific examples on $\mathsf{X}=\mathbb{R}^d$, 
we consider assumptions along the path 
$\pi(\lambda,dx)=\gamma(\lambda,x)dx/Z(\lambda)$ for $\lambda\in[0,1]$, 
where $\gamma(\lambda,x)=\gamma_0(x)^{1-\lambda}\gamma(x)^{\lambda}$ 
and $Z(\lambda)=\int_{\mathsf{X}}\gamma(\lambda,x)dx$. The densities 
$\gamma_0(x)$ and $\gamma(x)$ are assumed to be continuously differentiable.
We will 
write the expectation of $\varphi:\mathbb{R}^d\rightarrow\mathbb{R}$ with respect to 
$\pi(\lambda,dx)$ as $\pi(\lambda,\varphi)=\int_{\mathsf{X}}\varphi(x)\pi(\lambda,dx)$ 
and $\ell(x)=\log(\gamma(x)/\gamma_0(x))$.
\begin{assump}\label{assump:geo}
There exist constants $C,\zeta>0$ and a function $\beta:[0,1]\rightarrow\mathbb{R}_+$ with 
$\inf_{\lambda\in[0,1]}\beta(\lambda)>0$ such that for each $\lambda\in[0,1]$, 
the distribution $\pi(\lambda,dx)$ along the geometric path satisfies:
\begin{enumerate}[(i)]
	\item a Poincar\'{e} inequality with constant $\beta(\lambda)$, i.e. for all differentiable 
	$\varphi:\mathbb{R}^d\rightarrow\mathbb{R}$, we have 
	$\pi(\lambda,\varphi^2)-\pi(\lambda,\varphi)^2
	\leq\beta(\lambda)^{-1}\pi(\lambda, |\nabla\varphi|^2)$;
	
	\item a bound on the maximum log-likelihood, $\sup_{x\in\mathsf{X}}\ell(x) \leq Cd^{\zeta}$; 
	
	\item a bound on the expected log-likelihood, $\pi(\lambda,\ell) \geq -Cd^{\zeta}$;
	
	\item a bound on the expected squared norm of the log-likelihood, $\pi(\lambda,|\nabla\ell|^2) \leq Cd^{2\zeta}$. 
	
\end{enumerate}
\end{assump}

The Poincar\'{e} inequality is an isoperimetric condition 
with rich implications, 
such as the
exponential convergence of certain MCMC algorithms
\citep{vempala2019rapid} which can be used to
generalize the discussion in Section \ref{subsec:norm_const_var} by relaxing
the assumption of perfectly mixing kernels \citep{schweizer2012non}.  We refer
to references in \citet[p. 7 \& 16]{vempala2019rapid} for conditions to
verify a Poincar\'{e} inequality and we recall that it is implied by strong
log-concavity of the distribution in question.
Under Assumption \ref{assump:geo}, we can 
verify Assumptions \ref{assump:chisq} and \ref{assump:T}.  At step
$t\in[T]$, the $\chi^2$-divergence of 
$\pi_{t-1}(dx)=\pi(\lambda_{t-1},dx)$ from $\pi_{t}(dx)=\pi(\lambda_{t},dx)$ 
for $0\leq\lambda_{t-1}<\lambda_t\leq1$ can be bounded,
\begin{equation}\label{eqn:chisq_poincare}
	\chi^2(\pi_t|\pi_{t-1}) \leq \beta(\lambda_{t-1})^{-1}(\lambda_t-\lambda_{t-1})^2
	\int_{\mathsf{X}}\frac{\pi_t(x)}{\pi_{t-1}(x)}|\nabla\ell(x)|^2\pi_{t}(dx).
\end{equation}
This follows from Assumption \ref{assump:geo}(i) for the distribution $\pi(\lambda_{t-1},dx)$ 
and the function $\varphi(x)=\pi_t(x)/\pi_{t-1}(x)$. 
To upper bound the ratio of densities in \eqref{eqn:chisq_poincare}, we consider 
\begin{align}\label{eqn:logratiodensities}
	\log\pi_t(x)-\log\pi_{t-1}(x) &= (\lambda_t-\lambda_{t-1})\ell(x)-(\log Z_t-\log Z_{t-1}) \\
	&= (\lambda_t-\lambda_{t-1})(\ell(x)-\pi(\lambda_t^*,\ell))\notag
\end{align}
which holds for some $\lambda_t^*\in(\lambda_{t-1},\lambda_t)$ using the mean value theorem. 
Hence using Assumption \ref{assump:geo}(ii)-(iii), we have 
\begin{equation}
\sup_{x\in\mathsf{X}}\frac{\pi_t(x)}{\pi_{t-1}(x)} \leq \exp(2C(\lambda_t-\lambda_{t-1})d^{\zeta}).
\end{equation}
Applying this upper bound in \eqref{eqn:chisq_poincare}, Assumption \ref{assump:geo}(iv) and 
the lower bound $\underline{\beta}=\inf_{\lambda\in[0,1]}\beta(\lambda)$,
we obtain 
\begin{equation}
	\chi^2(\pi_t|\pi_{t-1}) \leq \underline{\beta}^{-1}(\lambda_t-\lambda_{t-1})^2\exp(2C(\lambda_t-\lambda_{t-1})d^{\zeta})Cd^{2\zeta}.
\end{equation}
If we construct a sequence $(\lambda_t)$ with increment $\lambda_t-\lambda_{t-1}=cd^{-\zeta}$, 
the constant $c>0$ can be chosen so that Assumption \ref{assump:chisq} holds, 
and Assumption \ref{assump:T} is satisfied since $T=O(d^{\zeta})$.

Formal studies 
on SMCS in high dimension include 
that of 
\citet{beskos2014stability}, which provides stability results in
settings where the target $\pi$ can be factorized into 
independent components, and discusses the behavior of the required number 
of bridging steps and of the effective sample size.
Relevant discussions can
also be found in Section 6 of \citet{schweizer2012non}, where solid
reasons are given to support a polynomial dimension dependence;
see also \citet{brosse2018normalizing}.
The appendices contain numerical illustrations of the performance
of SMCS in increasing dimensions.

\section{Parallel execution and confidence intervals}\label{sec:theory} 

Having specified the ingredients of SMCS,
the user has more than one way of running these algorithms,
which leads to different 
perspectives on the use of parallel processors
and on the quantification of errors. 

\subsection{Interacting particle systems}\label{subsec:bigsmc}

 SMCS 
are instances of interacting particle systems, 
or equivalently Monte Carlo approximations of Feynman--Kac models.  
This view has proven fruitful and allows the
application of readily-available results \citep{del2004feynman,del2013mean}.
In particular the estimators
$\pi_t^N(\varphi)$, for a function $\varphi$, and $Z_t^N$ satisfy central limit theorems:
\begin{align}
    \sqrt{N}(\pi_t^N(\varphi) - \pi_t(\varphi)) &\stackrel{d.}{\longrightarrow} \mathcal{N}(0,v_t(\varphi)),\label{eq:cltphi}\\
    \sqrt{N}(Z_t^N / Z_t - 1) & \stackrel{d.}{\longrightarrow} \mathcal{N}(0,v^\star_t), \label{eq:cltZ}
\end{align}
for each $t$ as $N\rightarrow\infty$, where $\stackrel{d.}{\longrightarrow}$ denotes 
convergence in distribution, and $v_t(\varphi),v_t^\star>0$ denote asymptotic variances.
Therefore valid confidence intervals can 
be derived from consistent estimators of the asymptotic variances.  
Such estimators 
were derived in
\citet{chan2013general,lee2015variance,du2021variance},
and address a long-standing gap on the quantification
of errors in SMC. 

We present a result from \citet{lee2015variance} that is 
valid when multinomial resampling is employed. 
We introduce the ``lineage'' of the $n$-th particle at step $t$: 
\begin{equation}
  \label{eq:lineage}
b_{t,t}^n = n, \quad \text{and} \quad b_{s-1,t}^n = a_{s-1}^{b_{s,t}^n} \quad \text{for}\; 1\leq s \leq t,
\end{equation}
where the ancestor indices $(a_t^n)$ are defined when resampling in Step 2(a) of Algorithm \ref{alg:smc}.
Since only the offspring of the particles indexed by $b_{0,t}^{1:N}$ survive 
at time $t$, we will refer to such indices as ``roots''. 
For a function $\varphi$, consider the quantity
\begin{align}
    V_t^N(\varphi) &= \pi_t^N(\varphi)^2 - 
    \left(\frac{N}{N-1}\right)^{t+1} \frac{1}{N^2} 
    \sum_{n,m\colon b_{0,t}^n \neq b_{0,t}^m} \varphi(x_t^n) \varphi(x_t^m),
    \label{eq:variancestimator}
\end{align}
which can be computed as a by-product of SMCS. 
Theorem 1 of \citet{lee2015variance}
states the convergence in probability of 
$N\cdot V_t^N(\varphi - \pi_t^N(\varphi))$ to $v_t(\varphi)$, and of 
$N\cdot V_t^N(1)$ to $v^\star_t$, as $N\to\infty$. 
We can directly write
\begin{align}
  V_t^N(1) &= 1 - \left(\frac{N}{N-1}\right)^{t+1} + \left(\frac{N}{N-1}\right)^{t+1}\frac{1}{N^{2}} \sum_{n\in[N]} 
  |\{m\colon b_{0,t}^m = b_{0,t}^n\}|^2.
  \label{eq:varianceestimatorZ}
\end{align}
The right-hand side features the cardinality of the set of
siblings of particle $n$, i.e. the particles that have the same ancestor at time zero.
If all particles were siblings, the sum would be of order $N^2$ and thus 
the estimated variance would be away from zero, but if all particles have a small number
of siblings, the variance estimator is of order $N^{-1}$.

The $N\to\infty$ regime underpinning the above results is compatible with
parallel computing since the propagation of particles and the calculation of
weights can be distributed across processors. The resampling step, on the other
hand, induces interactions and thus synchronization and communication
\citep{jun2012entangled,murray2016parallel}, so that SMCS are not
fully parallelizable.  Another limitation is that a direct implementation
requires $N$ particles in memory, which can be limiting in certain settings;
\citet{jun2014memory} propose a memory-efficient implementation.  Finally  SMCS
in the large $N$ regime are not ``anytime'': they run for $T$ steps before
returning their output and then stop.  In contrast, MCMC methods are easier to
interrupt and resume. These shortcomings have motivated variants where
individual particles can be added sequentially
\citep{brockwell2010sequentially,paige2014asynchronous,finke2018limit}. 

\subsection{Independent systems of fixed size}
\label{subsec:smallsmc}

Consider $R$ independent SMCS runs, with a fixed number of particles $N$.
The runs can be executed on parallel machines,
and denote by $(\pi^{N,r})_{r\in[R]}$ and by $(Z^{N,r})_{r\in[R]}$
the resulting particle approximations of $\pi$ and $Z$ respectively. 
We can obtain consistent approximations of $\pi$ and $Z$ as $R\to\infty$,
even though $N$ is fixed
\citep[e.g.][]{whiteley2016role, rainforth2016interacting}.
Since the normalizing constant estimator 
is unbiased, we can directly average $(Z^{N,r})_{r\in[R]}$ 
to obtain a consistent estimator of $Z$ as $R\to\infty$. 
Estimating expectations under $\pi$ seems more involved as 
the estimator $\pi_t^N(\varphi)$ is itself biased when $N$ is fixed.
We describe how to correct this in the framework 
of \citet{andrieu2010particle}. 

We select a particle among the $N$ available ones at the terminal step
of Algorithm \ref{alg:smc},
by sampling $k\in[N]$ with probabilities $w_T^{1:N}$,
and returning $x_T^{k}$. 
The distribution of all random variables generated by the procedure is 
\begin{align}
q^N(k, \bar{x}, \bar{a}) =
 \left\{\prod_{n\in[N]} \pi_0(x_0^n)\right\} \prod_{t=1}^T \left\{ r(a_{t-1}^{1:N}|w^{1:N}_{t-1}) \prod_{n\in[N]} M_t(x_{t-1}^{a_{t-1}^n},x_t^n)\right\}w_T^k,
\end{align}
where 
$\bar{x} = (x_t^{n})_{n\in[N]}$ for $0\leq t\leq T$ and 
$\bar{a} = (a_t^{n})_{n\in[N]}$ for $0\leq t \leq T-1$.
Next we define 
\begin{align}
	\label{eq:pitilde}
	\bar{\pi}^N(k, \bar{x}, \bar{a}) &= \frac{{Z}_T^N}{Z_T} q^N(k, \bar{x}, \bar{a}).
\end{align}
Under a mild assumption on the resampling scheme, \citet{andrieu2010particle} 
observe that \eqref{eq:pitilde} defines a valid probability distribution, and 
that its marginal distribution in $x_T^k$ is the target $\pi$. 
Therefore we can use $q^N$ as a proposal and $\bar{\pi}^N$ as a  target 
in an importance sampling argument, and the corresponding unnormalized weight is $Z_T^N$.   
For a function $\varphi$, a self-normalized importance sampling estimator after Rao-Blackwellizing 
the index $k$ is thus
\begin{align}
    \label{eq:combinationindepSMC} 
    \bar{\pi}^{R}(\varphi) &= \frac{\sum_{r\in[R]} Z^{N,r} \pi^{N,r}(\varphi)}{\sum_{r'\in[R]} Z^{N,r'}}, 
\end{align}
which approximates $\pi(\varphi)$ as $R\to\infty$, for any fixed $N$. Its 
asymptotic variance can be estimated consistently as $R\to\infty$ 
to construct confidence intervals \citep[Equation (9.8) in][]{owen_2013}.
There are practical benefits of the large $R$ asymptotics over the large $N$ asymptotics:
independent SMCS can be run on parallel machines without communication; 
results can be refined with more independent runs without hitting memory limits;
the procedure is simple to implement, to interrupt and to resume. 
More advanced schemes where ``islands'' of particles are allowed to communicate
have been studied in e.g. \citet{verge2015parallel,whiteley2016role}.

Equation \eqref{eq:pitilde} suggests the use of SMCS 
as an independent proposal in a Metropolis--Rosenbluth--Teller--Hastings algorithm
\citep{andrieu2010particle}.
Despite the iterative nature of MCMC, most of the computation here lies 
in the generation of the independent proposals, which can be fully parallelized. 
The approach lends itself to 
generic convergence diagnostics for MCMC \citep{brooks2011handbook},
and other tools developed for MCMC estimators, including 
variance reduction \citep{dellaportas2012control} and debiasing techniques \citep{middleton2019unbiased}.
The latter reference describes a generic strategy
that delivers unbiased estimators of $\pi(\varphi)$
using only standard SMCS runs, thus bypassing the design of algorithm-specific couplings 
as in \citet{jacob2020unbiased}. 

The performance of \eqref{eq:combinationindepSMC} clearly depends on the
performance of each run of SMCS. If $N$ is fixed to a low value, the
performance can still be satisfactory provided that the other algorithmic
ingredients are well-chosen. The appendices present numerical
experiments where the variance of $\log Z^N$ is seen to be stable in problems
of increasing dimension $d$ using a fixed value of $N$, thanks to an adequate
scaling of the number of intermediate steps.

\section{Illustrations} \label{sec:objects}

We illustrate some appealing properties of SMCS 
compared to MCMC methods 
in two simple examples; all details and the implementation
are described in appendices.
More challenging problems have been tackled with SMCS, e.g. in
Bayesian non-parametrics \citep{griffin2017sequential}, phylogenetic
inference \citep{wang2015jasa}, fiducial inference
\citep{cisewski2012generalized}, financial econometrics
\citep{fulop2013efficient,fulop2021bayesian}, large-scale graphical models
\citep{naesseth2014sequential}, partial differential
equations \citep{beskos2017multilevel} and experimental design
\citep{drovandi2014sequential,cuturi2020noisy}.

\subsection{Logistic regression\label{subsec:logistic}}

We consider a logistic regression 
$y=(y_1,\ldots,y_m)\in\left\{ 0,1\right\}^{m}$ on covariates
$x=(x_{1},\ldots,x_{m})\in\mathbb{R}^{m\times d}$. 
Under the model, $y_i$ is a Bernoulli variable with success probability 
$(1+\exp(-x_i^\top\beta))^{-1}$ where $\beta\in\mathbb{R}^d$ denote 
the regression coefficients. 
We use the ``forest cover type'' data
\citep{blackard2000comparison}, processed as in
\citet{collobert2002parallel}\footnote{\url{https://www.csie.ntu.edu.tw/~cjlin/libsvmtools/datasets/binary.html}.}.
The data contain cartographic information (relating to altitude, slope,
azimuth etc) for 30$m$ by 30$m$ cells in northern Colorado, along with the type
of cover (originally spruce/fir, lodgepole pine, Ponderosa pine,
cottonwood/willow, spruce/fir and aspen or Douglas-fir, and in
\citet{collobert2002parallel} this was simplified to lodgepole pine
versus the other categories combined).  With a logistic regression, 
we predict the cover type using cartographic variables. 
There are $d=11$ regression coefficients including the intercept,
and the prior is Normal$(0,10)$ on each coefficient unless specified otherwise. 

We illustrate the sequential aspect of Bayesian updating
with SMCS using the path of partial posteriors; 
other paths are considered in the appendices. 
Figures \ref{fig:merging:beta1}-\ref{fig:merging:beta8} show
a phenomenon called ``merging''  
whereby posteriors resulting from different
priors eventually coincide as more observations are introduced. 
We observe that certain components of the posterior 
``merge'' faster than others.
Similar figures could help to visualize the 
Bernstein-von Mises phenomenon, whereby the posterior distribution becomes
closer to a Normal distribution as more data get assimilated. 

\begin{figure*}[t!]
    \centering
    \begin{subfigure}[t]{0.45\textwidth}
        \centering
        \includegraphics[width=\textwidth]{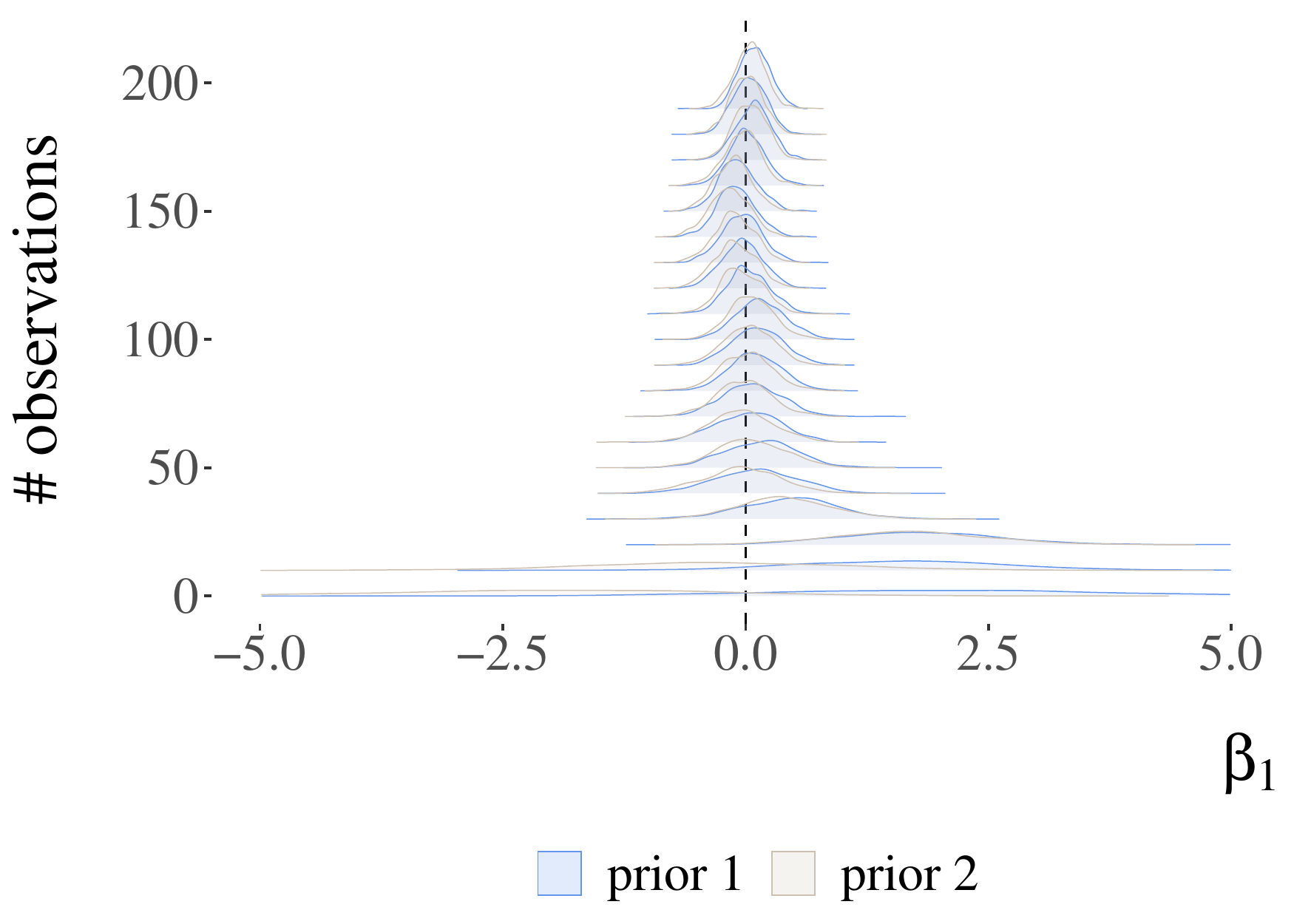}
        \caption{\label{fig:merging:beta1} Marginal on $\beta_1$.}
    \end{subfigure}
    ~
    \begin{subfigure}[t]{0.45\textwidth}
        \centering
        \includegraphics[width=\textwidth]{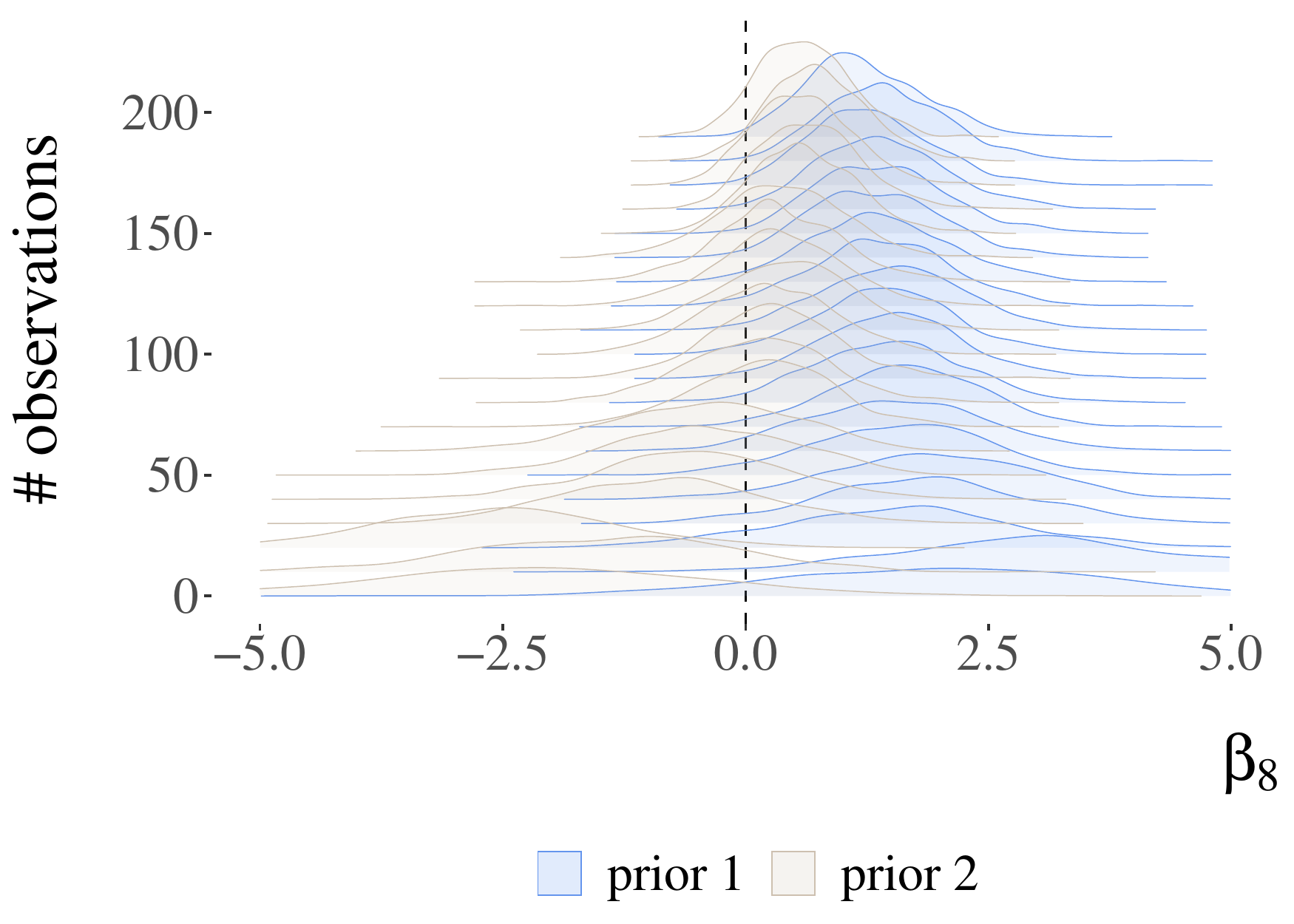}
        \caption{\label{fig:merging:beta8} Marginal on $\beta_8$. }
    \end{subfigure}
    ~
    \begin{subfigure}[t]{0.45\textwidth}
        \centering
        \includegraphics[width=\textwidth]{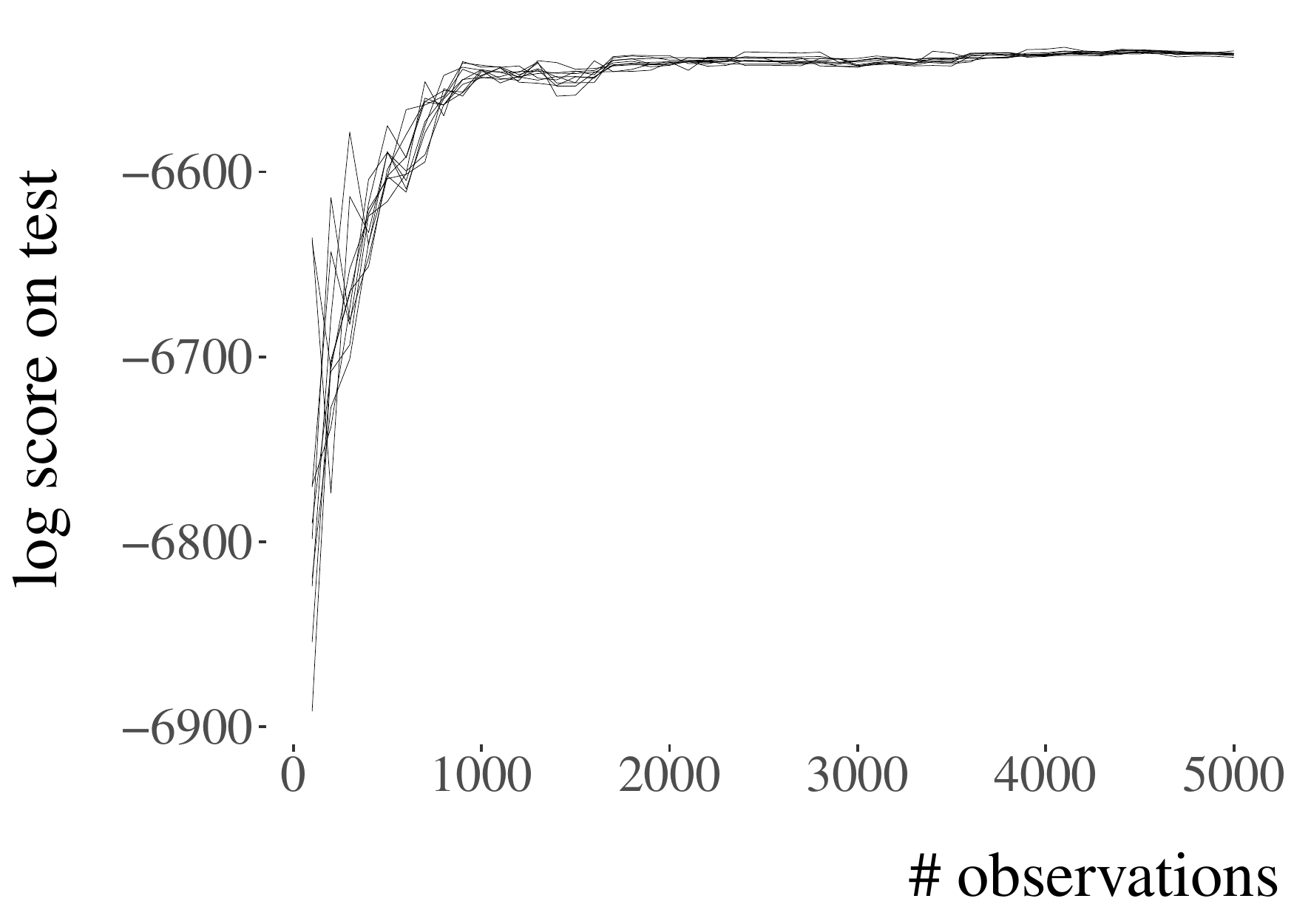}
        \caption{\label{fig:pred1} Prediction of first 5000 observations.}
    \end{subfigure}
    ~
    \begin{subfigure}[t]{0.45\textwidth}
        \centering
        \includegraphics[width=\textwidth]{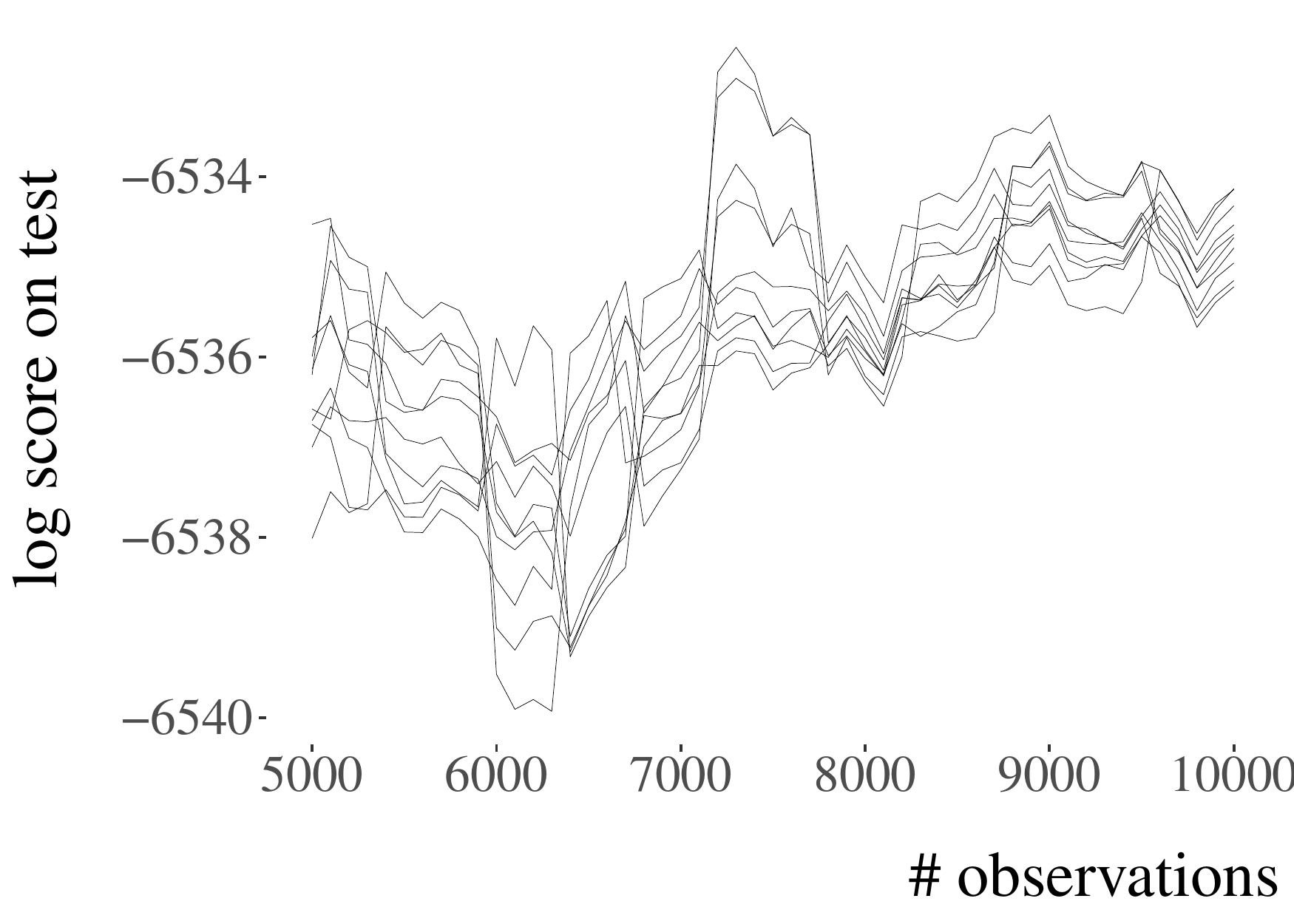}
        \caption{\label{fig:pred2} Prediction of next 5000 observations.}
    \end{subfigure}
    
    \caption{\label{fig:merging:pred} Logistic regression with forest cover type data. Evolution
    of the posterior distribution of $\beta_1$ (\emph{top-left}) and $\beta_8$ (\emph{top-right}) 
    as more data are assimilated, with initialization from the priors $\mathcal{N}(2,3)$ 
    (\emph{blue}) and $\mathcal{N}(-2,3)$ (\emph{beige}).
  Performance
  of the posterior predictive distribution on a test data set as the first 5000 (\emph{bottom-left}) 
  and next 5000 (\emph{bottom-right}) observations are assimilated, estimated using independent runs of SMCS.}
\end{figure*}

Sequential inference allows us to monitor not only the evolution of beliefs 
but also measures of predictive performance. For example, Figures
\ref{fig:pred1}-\ref{fig:pred2} show the logarithmic score associated with the
posterior predictive distribution as the number of observations increases,
on a test data set.  Predictive performance increases significantly
as we start to assimilate data.  After a certain point the predictive
performance seems to stagnate.  Indeed, under model misspecification, there is
no guarantee that the posterior predictive performance would improve
with more data.  The ability to monitor performance can be helpful when
deciding whether the model under consideration is able to benefit from the
inclusion of more data. The arbitrariness of the ordering of the observations
in the setting of regression can be addressed by averaging over orderings, as
described in the appendices, where it is also shown how Bayesian
asymptotics provide efficient strategies for initializing SMCS. 

\subsection{Susceptible-Infected-Recovered model\label{subsec:SIR}}

\begin{figure*}[t!]
    \centering
    \begin{subfigure}[t]{0.45\textwidth}
        \centering
        \includegraphics[width=\textwidth]{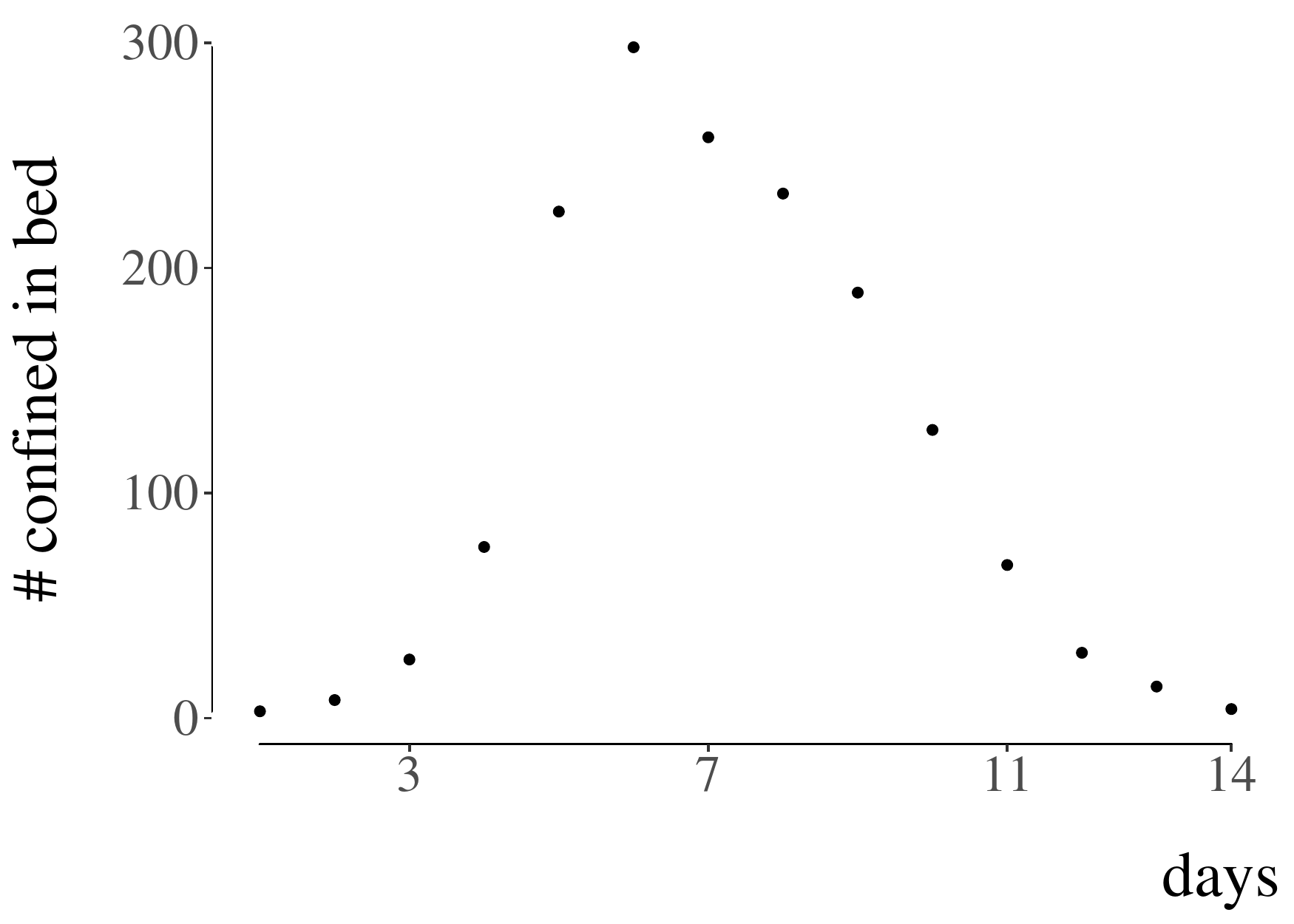}
        \caption{\label{fig:boardingschool:obs} Observations.}
    \end{subfigure}
    ~
    \begin{subfigure}[t]{0.45\textwidth}
        \centering
        \includegraphics[width=\textwidth]{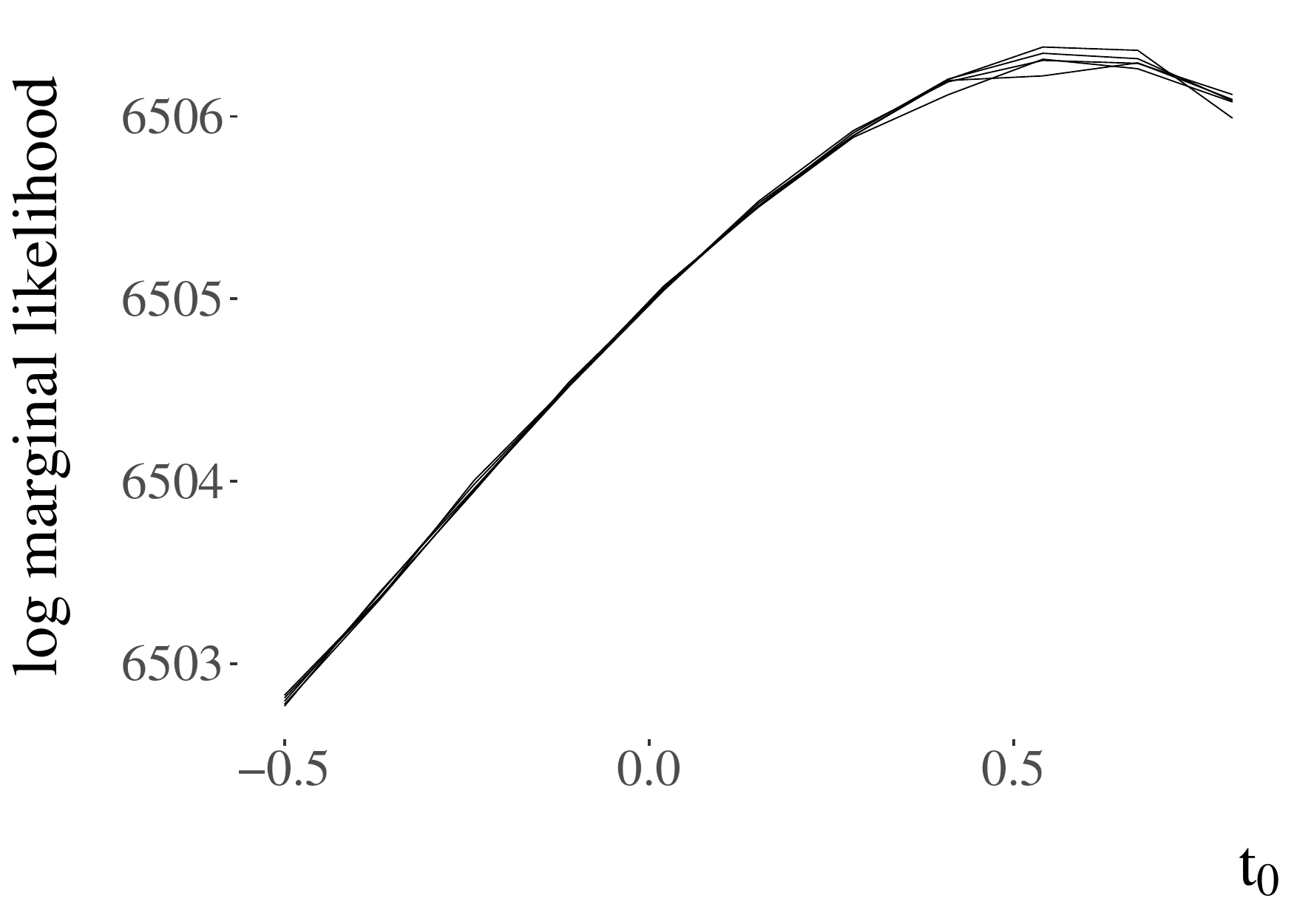}
        \caption{\label{fig:boardingschool:logz} Log-marginal likelihood.}
    \end{subfigure}
    ~
    \begin{subfigure}[t]{0.45\textwidth}
        \centering
        \includegraphics[width=\textwidth]{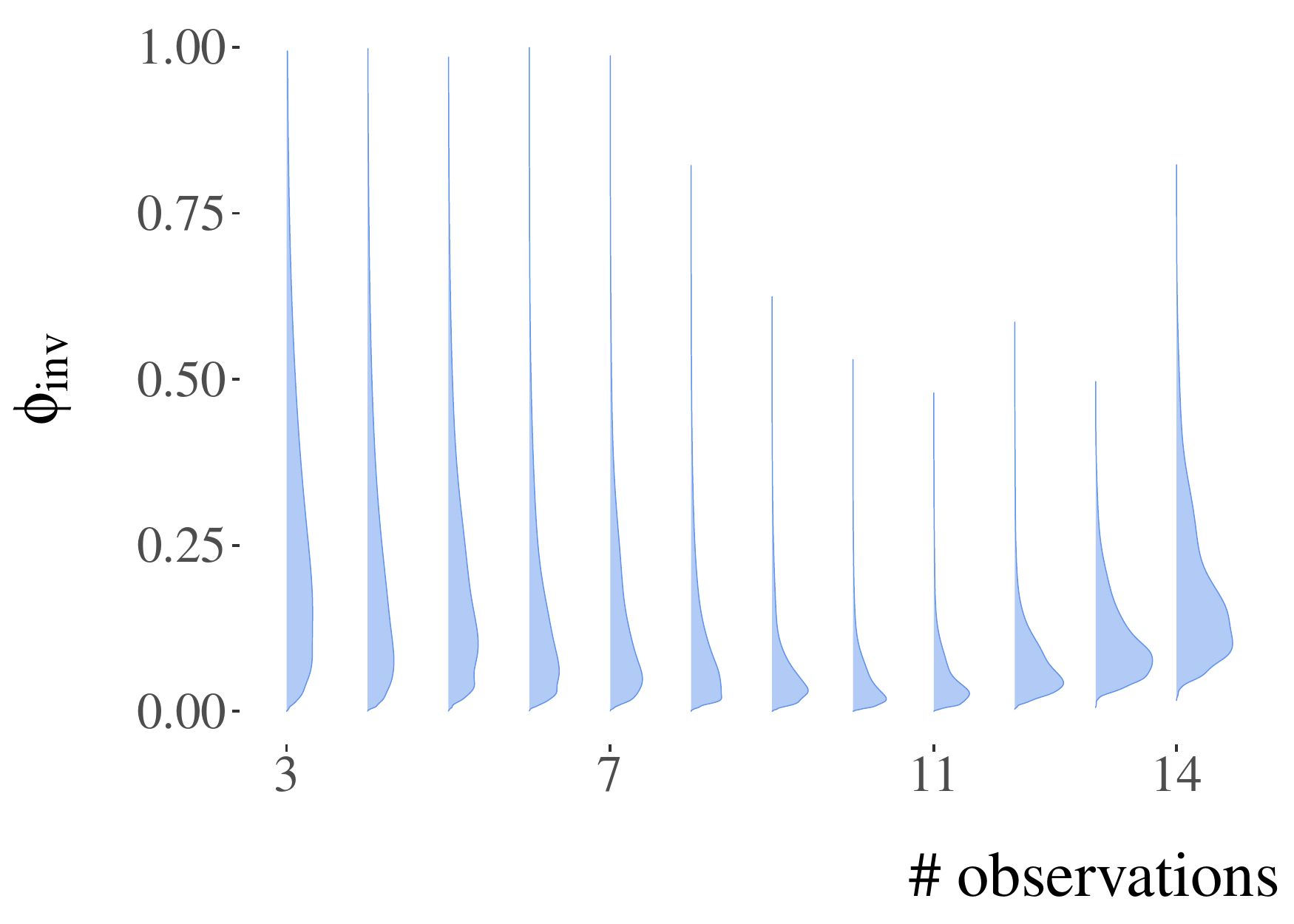}
        \caption{\label{fig:boardingschool:phiinv} Posterior of $\phi_{\text{inv}}$.}
    \end{subfigure}
    ~
    \begin{subfigure}[t]{0.45\textwidth}
        \centering
        \includegraphics[width=\textwidth]{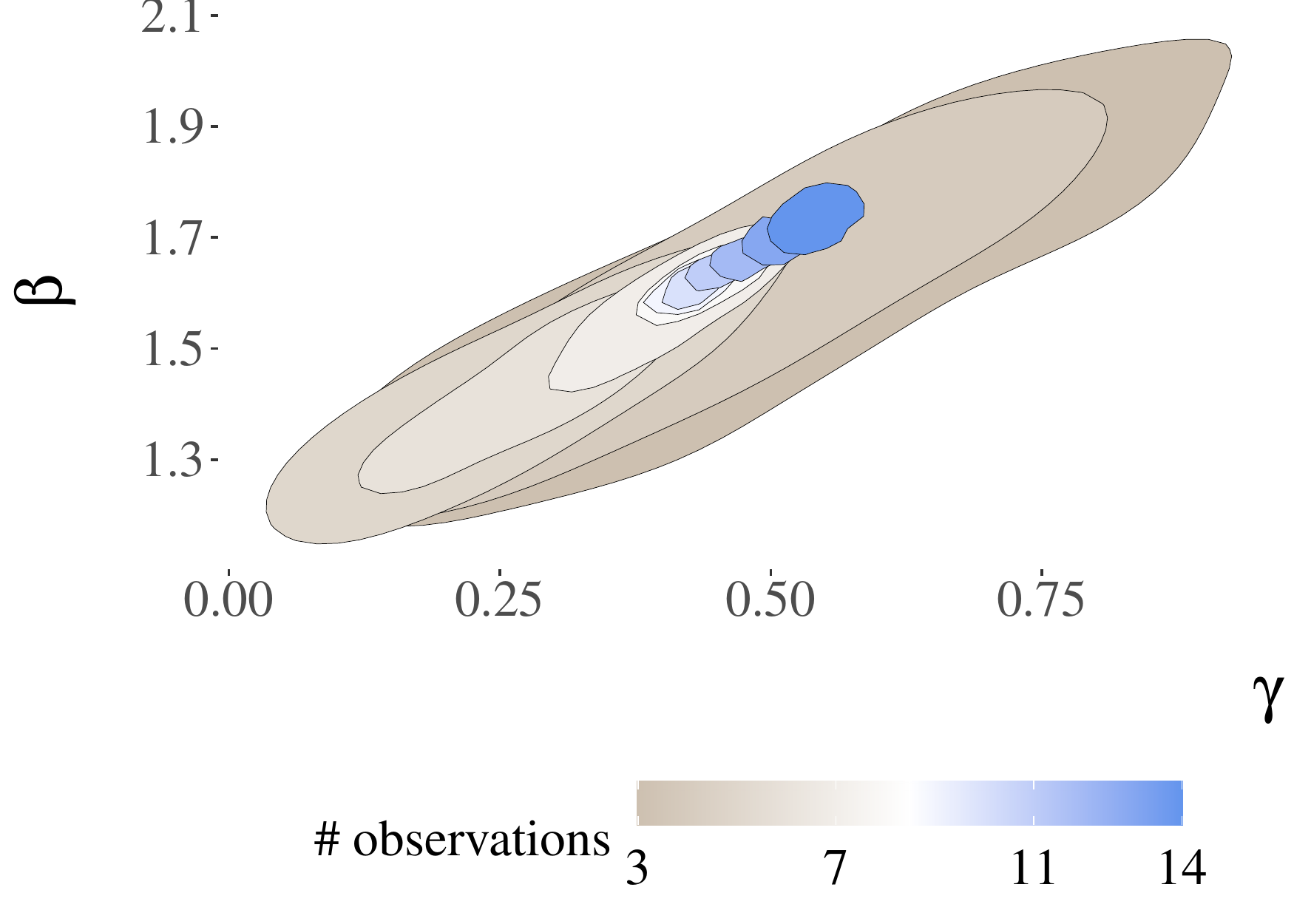}
        \caption{\label{fig:boardingschool:contours} Posterior of $(\gamma,\beta)$.}
    \end{subfigure}
    \caption{\label{fig:boardingschool} SIR model with boarding school data. 
    Observations of daily counts (\emph{top-left}). Log-marginal likelihood of 
    the initial time $t_0$ at which the first individual is assumed to be infected (\emph{top-right}). 
    Evolution of the marginal posterior distribution of $\phi_{\text{inv}}$ (\emph{bottom-left}) and 
    of $(\gamma,\beta)$ (\emph{bottom-right}) as more data are assimilated.}
\end{figure*}

Another setting where sequential inference is particularly relevant is the
modeling of disease outbreaks. Parameter calibration involves blending prior
knowledge with data arriving regularly, typically daily or weekly. 
We consider a simple deterministic Susceptible-Infected-Recovered (SIR) 
model \citep{bacaer2012model}.
Inference for such models can be done with MCMC \citep{grinsztajn2020bayesian}. 
We consider an example from that article, 
using the classical boarding school data of 
daily counts of pupils confined to bed during an influenza outbreak,
shown in Figure \ref{fig:boardingschool:obs}. The model is described by the differential equations
\begin{align}
  \label{eq:sirode}
  \frac{dS}{dt} &= - \beta S I / n,\quad 
  \frac{dI}{dt} = \beta S I / n - \gamma I, \quad
  \frac{dR}{dt} = \gamma I,
\end{align}
where $n=763$ is the total number of school children, $S$, $I$ and $R$
represent the number of susceptible, infected and recovered children,
respectively, and $\gamma,\beta>0$ are parameters to be inferred.  We assume an
initial condition of $(S,I,R)=(n-1,1,0)$ at time $t_0=0$, i.e. with an infected
individual.  The observations, which begin at time $t=1$, are assumed to be
noisy measurements of the number $I$ of infected children that day. The
observation noise is modeled as a Negative Binomial distribution parametrized
by $\phi_{\text{inv}}>0$.  Priors on $\gamma,\beta,\phi_{\text{inv}}$ are taken 
as $\mathcal{N}(0.4,0.5^2)$, $\mathcal{N}(2,1)$ (truncated to $\mathbb{R}_+$) and an
exponential distribution with rate 5, respectively.  The Stan implementation in
\citet{grinsztajn2020bayesian} provides a function that evaluates the posterior
log-density, which we use in an adaptive SMCS for the path of partial posteriors. 

The bottom row of Figure
\ref{fig:boardingschool} displays the time evolution of the posterior
distribution of parameters.  These types of visualization could be used, for
example, to study how many observations are necessary to obtain a desired
precision on the parameter estimates.  Lastly, we consider a simple procedure
to infer the initial time $t_0$ at which the first individual is assumed to be
infected.  Figure \ref{fig:boardingschool:logz} plots the marginal likelihood
of $t_0$, which is the normalizing constant of the corresponding posterior
distribution, obtained here by running SMCS 10 times independently for
different values of $t_0$; we observe a peak around $t_0 = 0.5$.

\section{Discussion} \label{sec:discussion}

Consider a standard MCMC setting, where we run $R$ chains independently for $T$ steps each, 
possibly after some early adaptive phase and discarding the first samples as burn-in. 
From this familiar situation, we might want to:
  1) parallelize computation across chains,
    ideally with a large $R$ 
    and a small $T$;
  2) allow the $R$ chains to communicate 
  in order to accelerate their
    exploration of the state space;
  3) use different Markov kernels depending on the marginal
    distribution of the chains at the current iteration;
  4) estimate the normalizing constant;
  5) approximate not a single but multiple, related target distributions.
There are various ways of addressing any of these points, but SMCS provide a
unified and principled strategy to address them all.  We can add to this list:
for example there are documented advantages of particle methods over Markov
chains for multimodal target distributions
\citep{schweizer_2012_multimodal,paulin2019error},
and unbiased estimators of normalizing constants lead to
useful evidence lower bounds for variational inference 
\citep{naesseth2018}. The most important message
is perhaps that SMCS provide a viable alternative to MCMC with distinct advantages 
that can help statisticians.

We take a cautious view regarding performance comparisons between SMCS
and MCMC algorithms. There are many tuning choices involved in both 
families of algorithms, thus one should not 
expect to draw fully general conclusions about one algorithm being superior to another. Comparisons
can be informative in specific cases
\citep[e.g.][]{matthews2022continual}. More often 
comparisons are made between SMCS and annealed importance sampling \citep[e.g.][]{heng2021gibbs} 
or between variants of SMCS \citep[e.g.][]{salomone2018unbiased}. 
In principle, any efficient MCMC
algorithm could also be used as an ingredient in SMCS, but the choice of paths
might not be obvious.  For example when the target distribution is supported on
a manifold \citep[see e.g.][]{diaconis2013sampling}, it might be difficult to define a
suitable initial distribution for SMCS, while any point on the manifold provides
a valid start for MCMC.  Also, as of today the literature on
convergence diagnostics is much more developed for MCMC methods \citep{roy2020convergence} than for SMCS.
Methods discussed in Section \ref{subsec:bigsmc} to construct valid confidence
intervals for SMCS estimators are only recent, while the construction
of confidence intervals using independent SMCS runs 
as in Section \ref{subsec:smallsmc} seems to be rarely employed.

Why are sequential Monte Carlo samplers not used more often?  The flexibility in the choice  of
paths and Markov kernels may appear overwhelming to new users. 
Despite useful efforts
to automatize the design of SMCS, for example 
using stochastic optimization \citep{fearnhead2013adaptive},
or Generative Adversarial Networks \citep{kempinska2017adversarial},
there remains a number of tuning choices to be addressed in any 
specific case, which may be a barrier even to computationally-minded statisticians.
In addition, software implementations of SMCS exist
\citep{wood2014new,salvatier2016probabilistic,murray2018automated} but are not
as widely used as MCMC software such as Stan
\citep{carpenter2017stan} and do not benefit from a comparable community support.
We have described reasons for SMCS to be implemented more often
by statisticians in the future.  

The code used to generate the figures of this 
article is available at \url{https://github.com/pierrejacob/smcsamplers}.

\paragraph*{Acknowledgements}

This work was funded by CY Initiative of Excellence (grant ``Investissements
d'Avenir'' ANR-16-IDEX-0008).  Pierre E. Jacob gratefully acknowledges support
by the National Science Foundation through grants DMS-1712872 and DMS-1844695.
The authors thank Elena Bortolato, Nicolas Chopin, Arnaud Doucet, Sam Power,
Rob Salomone, Leah South, Ian and Elizabeth Taylor for useful discussions.

\bibliographystyle{agsm}
\bibliography{Tempering.bib}

\appendix

\section{Particle filters and SMC samplers} 

\paragraph{General SMC method.}
To elucidate the connections between particle filters and SMCS, we show here that these two algorithms are specific cases of a general SMC method that approximates a sequence of target distributions $(\tilde{\pi}_t)$. 
Each target distribution 
\begin{align}\label{eqn:gsmc_target}
	\tilde{\pi}_t(dx_{0:t})=\tilde{\gamma}_t(x_{0:t})dx_{0:t}/\tilde{Z}_t
\end{align}
is defined on the product space $(\mathsf{X}^{t+1}, \mathscr{X}^{t+1})$, where $\tilde{\gamma}_t(x_{0:t})$ is an unnormalized density and $\tilde{Z}_t=\int_{\mathsf{X}^{t+1}}\tilde{\gamma}_t(x_{0:t})dx_{0:t}$ is a normalizing constant (with $\tilde{Z}_0=1$). 

The general SMC method described in Algorithm \ref{alg:gsmc} combines sequential importance sampling and resampling. As input, it requires a sequence of proposal kernels $(q_t)$ on $(\mathsf{X}, \mathscr{X})$. At step $t$, this defines the proposal distribution 
\begin{align}
	\tilde{q}_t(dx_{0:t}) = \tilde{\pi}_0(dx_0)\prod_{s=1}^tq_s(x_{s-1},dx_s).
\end{align}
The weight function can be written as
\begin{align}
	\tilde{w}_t(x_{0:t}) = \tilde{\gamma}_t(x_{0:t})/\tilde{q}_t(x_{0:t}) = \prod_{s=1}^tw_s(x_{0:s}),
\end{align}
where the incremental weight function is 
\begin{align}\label{eqn:gsmc_incremental}
	w_t(x_{0:t}) = \frac{\tilde{\gamma}_t(x_{0:t})}{\tilde{\gamma}_{t-1}(x_{0:t-1})q_t(x_{t-1},x_t)}.
\end{align}
Representing the target distribution as 
\begin{align}
	\tilde{\pi}_t(dx_{0:t}) = \prod_{s=1}^tw_s(x_{0:s})\tilde{q}_t(dx_{0:t})/\tilde{Z}_t
\end{align}
allows us to adopt the Feynman--Kac formalism introduced by
\citet{del2004feynman,del2013mean}, where $(w_t)$ are seen as potential
functions that reassign the probability mass of $\tilde{q}_t$ to obtain
$\tilde{\pi}_t$, and the marginal distributions
$\tilde{\pi}_t(dx_{t})=\int_{\mathsf{X}^t}\tilde{\pi}_t(dx_{0:t})$ are referred
to as (updated) Feynman--Kac models.  As output, the algorithm returns weighted
particles $(w_t^n,x_{0:t}^n)_{n\in[N]}$ approximating $\tilde{\pi}_t$ as
$N\rightarrow\infty$, and an unbiased estimator $\tilde{Z}^N_t$ of
$\tilde{Z}_t$ which is consistent as $N\rightarrow\infty$.

\begin{algorithm}
	\caption{General sequential Monte Carlo method}\label{alg:gsmc}
\begin{flushleft}
	\textbf{Input:} 
	sequence of distributions $(\tilde{\pi}_t)$, proposal Markov kernels $(q_t)$, 
	resampling distribution $r(\cdot|w^{1:N})$ on $[N]^N$ where $w^{1:N}$ is an $N$-vector of probabilities.
\begin{enumerate}
\item Initialization.
\begin{enumerate}
	\item Sample particle $x_0^n$ from $\tilde{\pi}_0(\cdot)$ for $n\in[N]$ independently.
	\item Set $w_0^n = N^{-1}$ for $n\in[N]$.
\end{enumerate}
\item For $t \in [T]$, iterate the following steps.
\begin{enumerate}
	\item Sample ancestor indices $(a_{t-1}^{n})_{n\in[N]}$ from $r(\cdot|w^{1:N}_{t-1})$,
	 \item[] and define $\check{x}_{0:t-1}^n = x_{0:t-1}^{a_{t-1}^n}$ for $n\in[N]$.
    \item Sample particle $x_t^n\sim q_t(\check{x}_{t - 1}^n,\cdot)$ and set $x_{0:t}^n=(\check{x}_{0:t-1}^n,x_t^n)$ for $n\in[N]$.
	\item Compute weights $w_t(x_{0:t}^n)$ for $n\in[N]$, and set $w_t^n \propto w_t(x_{0:t}^n)$ such that $\sum_{n\in[N]} w_t^n = 1$.
\end{enumerate}
\end{enumerate}
\textbf{Output:} weighted particles $(w_t^n,x_{0:t}^n)_{n\in[N]}$ approximating $\tilde{\pi}_t$,
and estimator $\tilde{Z}^N_t = \prod_{s=1}^t N^{-1} \sum_{n\in[N]} w_s(x_{0:s}^n)$ of $\tilde{Z}_t$ for $t\in[T]$.
\end{flushleft}
\end{algorithm}

\paragraph{Particle filters.} 
We now introduce state space models which are also known as hidden Markov models. 
Consider a latent Markov chain $(x_t)_{t\geq0}$ defined on $(\mathsf{X}, \mathscr{X})$, initialized as $x_0\sim\pi_0$ and evolving for each time step $t\geq 1$ according to a Markov kernel $f$, i.e. $x_t|x_{t-1}\sim f(x_{t-1},\cdot)$. We assume access to $\mathsf{Y}$-valued observations $(y_t)_{t\geq1}$ that are modelled as conditionally independent given $(x_t)_{t\geq0}$, with observation density $g$ on $(\mathsf{Y}, \mathscr{Y})$, i.e. $y_t|x_t\sim g(x_t,\cdot)$. 

Given observations collected up to time $t$, sequential state inference is based on the posterior distribution 
\begin{align}\label{eqn:posterior_filter}
	p(dx_{0:t}|y_{1:t}) = \frac{p(dx_{0:t})p(y_{1:t}|x_{0:t})}{p(y_{1:t})},
\end{align}
where the joint distribution of the states is $p(dx_{0:t}) = \pi_0(dx_0)\prod_{s=1}^tf(x_{s-1},dx_s)$ and 
the conditional likelihood of the observations is $p(y_{1:t}|x_{0:t}) = \prod_{s=1}^tg(x_s,y_s)$. We will also be interested in the marginal likelihood $p(y_{1:t})=\int_{\mathsf{X}^{t+1}}p(dx_{0:t},y_{1:t})$ when there are unknown parameters in the model to be inferred. From \eqref{eqn:posterior_filter}, we can derive other quantities of interest such as the filtering distribution $p(dx_t|y_{1:t})$, defined as the last marginal of $p(dx_{0:t}|y_{1:t})$, and the state predictive distribution $p(dx_{t+1}|y_{1:t})=\int_{\mathsf{X}}f(x_t,dx_{t+1})p(dx_t|y_{1:t})$. 
Particle filters can be understood as specific cases of SMC methods to sequentially approximate the posterior distribution $\tilde{\pi}_t(dx_{0:t})=p(dx_{0:t}|y_{1:t})$ (with $\tilde{\pi}_0=\pi_0$) and the marginal likelihood $\tilde{Z}_t=p(y_{1:t})$. In this setting, the incremental weight function in \eqref{eqn:gsmc_incremental} reduces to 
\begin{align}
	w_t(x_{t-1},x_t) = \frac{f(x_{t-1},x_t)g(x_t,y_t)}{q_t(x_{t-1},x_t)}.
\end{align}
Different choices of proposal kernels $(q_t)$ give rise to distinct SMC methods. For example, the bootstrap particle filter of \citet{gordon1993novel} corresponds to Algorithm \ref{alg:gsmc} with $q_t(x_{t-1},dx_t)=f(x_{t-1},dx_t)$ and $w_t(x_t)=g(x_t,y_t)$ for all $t$. 

\paragraph{SMC samplers.}
We now cast SMCS presented in this article as specific cases of SMC methods. 
Given a sequence of target distributions $(\pi_t)$ and backward Markov kernels $(L_t)$ on $(\mathsf{X}, \mathscr{X})$, the 
target distribution in \eqref{eqn:gsmc_target} is 
\begin{align}\label{eqn:extendedtarget_smcsamplers}
	\tilde{\pi}_t(dx_{0:t}) = \pi_t(dx_t)\prod_{s=1}^{t}L_{s-1}(x_s,dx_{s-1}),
\end{align}
with $\tilde{\pi}_0=\pi_0$. Note that \eqref{eqn:extendedtarget_smcsamplers} has $\pi_t$ as the marginal distribution on $x_t$ and the normalizing constant is $\tilde{Z}_t=Z_t$. In this case, the proposal kernels $(q_t)$ correspond to the forward kernels $(M_t)$ defined on $(\mathsf{X}, \mathscr{X})$ and the incremental weight function \eqref{eqn:gsmc_incremental} reduces to the weight function in \eqref{eq:smc-incremental-weight}. 

Under these settings, one can check that Algorithm \ref{alg:gsmc} recovers the generic SMCS described in Algorithm \ref{alg:smc}, with the exception that we only keep track of the particles approximating the target distribution at each step. The latter is sufficient due to the simplified form of the weight function \eqref{eq:smc-incremental-weight} and the fact that only the terminal time marginal distribution of \eqref{eqn:extendedtarget_smcsamplers} is of interest. 

In summary, we see that particle filters and SMCS are instances of SMC methods that approximate different sequences of target distributions $(\tilde{\pi}_t)$, defined by either the specificity of the problem in \eqref{eqn:posterior_filter} or algorithmic choices in \eqref{eqn:extendedtarget_smcsamplers} via the specification of backward kernels. The terminal time marginal distribution $\tilde{\pi}_t(dx_{t})$ and normalizing constant $\tilde{Z}_t$ represent the filtering distribution $p(dx_t|y_{1:t})$ and marginal likelihood $p(y_{1:t})$ in particle filtering, and a target distribution $\pi_t(dx_t)$ and its associated normalizing constant $Z_t$ for SMCS. 
The proposal kernel $q_t$ corresponds to the state transition $f$ in the case of a bootstrap particle filter and the forward Markov kernel $M_t$ in SMCS.

\section{Unadjusted Hamiltonian Monte Carlo moves}

As an alternative to the unadjusted Langevin moves described
in Section \ref{subsec:choosemarkov},
we can consider kernels constructed using Hamiltonian dynamics 
\citep{duane1987hybrid} that target 
$\tilde{\pi}_t(dx_t,dv_t)=\pi_t(dx_t)\mathcal{N}(dv_t;0,\Omega)$ for $(x_t,v_t)\in\mathbb{R}^d\times\mathbb{R}^d$. 
Here $x_t$ are the original states, $v_t$ are auxiliary variables and 
$\Omega\in\mathbb{R}^{d\times d}$ denotes a ``mass matrix''.
Given a sample $x_{t-1}$ from $\pi_{t-1}$ at step $t-1$, 
we sample $v_{t-1}\sim \mathcal{N}(0, \Omega)$, so that the pair 
$(x_{t-1}, v_{t-1})$ follows $\tilde{\pi}_{t-1}$. 
We define the initial position $q(0)=x_{t-1}$ 
and initial momentum $p(0) = v_{t-1}$ of a fictitious object undergoing 
Hamiltonian dynamics, with Hamiltonian function
$H_t(q,p)=-\log\pi_t(q) + p^\top\Omega^{-1}p/2$. 
The associated dynamics is commonly discretized using 
the leap-frog integrator,
with a step size $\varepsilon>0$ and
a number of steps $m\in\mathbb{N}$, yielding a trajectory $(q(\ell),p(\ell))$ for $\ell=1,\ldots,m$.
Finally, we set $x_t=q(m)$
and $v_t=p(m)$.  We write the composition of leap-frog iterations as
$\Phi_t^\ell(q(0),p(0))=(q(\ell),p(\ell))$ for $\ell\in[m]$. 
The transition from $(x_{t-1},v_{t-1})$ to $(x_{t},v_{t})$ defines a deterministic forward 
kernel $M_t$, namely a Dirac mass on $\Phi_t^m(x_{t-1},v_{t-1})$. 

As the Hamiltonian is not conserved exactly under time-discretization, $M_t$ is
not $\tilde{\pi}_t$-invariant.  
It is again possible to correct the
discretization error using importance sampling to target $\tilde{\pi}_t(dx_t,dv_t)$  
with proposal $q_t(dx_t,dv_t)=(\tilde{\pi}_{t-1}\#\Phi_t^m)(dx_t,dv_t)$.
The $\#$ notation refers to 
the push-forward operator,
so that $(\tilde{\pi}_{t-1}\#\Phi_t^m)$
is the measure obtained by sampling from $\tilde{\pi}_{t-1}$ and applying the map $\Phi_t^m$. 
Using reversibility and volume preservation properties of $\Phi_t^m$, the proposal density can be computed using change of
variables, i.e. $q_t(x_t,v_t)=\tilde{\pi}_{t-1}(x_{t-1},v_{t-1})$ where
$(x_{t-1},v_{t-1})=(\Phi_t^m)^{-1}(x_t,v_t)$ is obtained using the inverse map.
The resulting importance weight is 
\begin{align}\label{eqn:HIS_weights}
	w_t(x_{t-1},v_{t-1},x_t,v_t) \propto \frac{\tilde{\pi}_t(x_t,v_t)}{\tilde{\pi}_{t-1}(x_{t-1},v_{t-1})} 
	= \frac{\exp(-H_t(x_t,v_t))}{\exp(-H_{t-1}(x_{t-1},v_{t-1}))},
\end{align}
which corresponds to the choice of backward kernel 
\begin{align}
L_{t-1}((x_t,v_t),dx_{t-1},dv_{t-1}) 
=\delta_{(\Phi_t^m)^{-1}(x_t,v_t)}(dx_{t-1},dv_{t-1}). 
\end{align}
The above arguments and related ideas can be found in \citet{jarzynski2000hamiltonian,neal2005hamiltonian,scholl2006proof}, 
and have been recently employed in variational inference frameworks to 
tune algorithmic parameters \citep{geffner2021mcmc,zhang2021differentiable}. 

We now discuss some extensions of the above framework. 
Firstly, one could replace $\Phi_t^m$ with other reversible and volume-preserving maps.
Secondly, we can consider several iterations of momentum refreshment 
and leap-frog integration, i.e. initializing at $x_{t,0}=x_{t-1}$, we would sample
$\tilde{v}_{t,i-1}\sim\mathcal{N}(0,\Omega)$ and set
$(x_{t,i},v_{t,i})=\Phi_t^m(x_{t,i-1},\tilde{v}_{t,i-1})$ for $i\in[I]$. 
One could also benefit from partial momentum refreshment \citep{horowitz1991generalized} by updating $\tilde{v}_{t,i}$ with 
an autoregressive process that leaves $\mathcal{N}(0,\Omega)$ invariant. 
In contrast to compositions of $\pi_t$-invariant kernels that do not affect
importance weights, we have to modify \eqref{eqn:HIS_weights} to account for
the additional iterations. 
Thirdly, in the spirit of the work by \citet{neal1994improved,nishimura2018recycling} 
for HMC and \citet{wastefreeSMC} for SMC, we can use all iterates in the leap-frog 
integrator instead of only the terminal ones,
by considering all the proposals
$q_t^{\ell}(dx_t,dv_t)=(\tilde{\pi}_{t-1}\#\Phi_t^\ell)(dx_t,dv_t)$ for all
$\ell\in[m]$ when forming an importance sampling approximation of
$\tilde{\pi}_t(dx_t,dv_t)$. In Algorithm \ref{alg:smc}, one would have $N\times
m$ instead of $N$ samples to consider in Steps 2(b) and 2(c); the resampling
operation in Step 2(a) would then select $N$ particles among the $N\times m$
weighted samples.  
Since
the use of multiple proposals within importance sampling is consistent in the
limit of the number of samples, it follows that the resulting SMCS will
also be consistent as $N\rightarrow\infty$.

\section{Numerical experiments on Gaussians \label{subsec:experiments}}

We consider numerical experiments on Normal distributions in varying dimensions $d$. 
We set $\pi_0(dx)=\mathcal{N}(x;\mu_0,\Sigma_0)dx$ 
with $\mu_0=(1,\ldots,1)$, $\Sigma_0=\textrm{diag}(0.5,\ldots,0.5)$ 
and $\pi(dx)=\mathcal{N}(x;\mu,\Sigma)dx$ 
with $\mu=(0,\ldots,0)$, $\Sigma=\textrm{diag}(1,\ldots,1)$. 
Despite the simple setup, standard importance sampling 
would give rise to estimators with infinite variance.  
We consider a geometric path 
of distributions, all of which are Normal. 
For each $t\in[T]$, we employ a $\pi_t$-invariant HMC kernel for $M_t$, with step size 
$\varepsilon = d^{-1/4}$ and $m=\lceil d^{1/4}\rceil$ leap-frog steps.
The ``mass matrix'' $\Omega$ is taken as diagonal and adapted 
using the empirical marginal precisions computed from the 
particle approximations. The backward kernel $L_{t-1}$ is taken as the time reversal of $M_t$. 
All simulations employ multinomial resampling. 

\begin{figure*}[t!]
    \centering
    \begin{subfigure}[t]{0.45\textwidth}
        \centering
        \includegraphics[width=\textwidth]{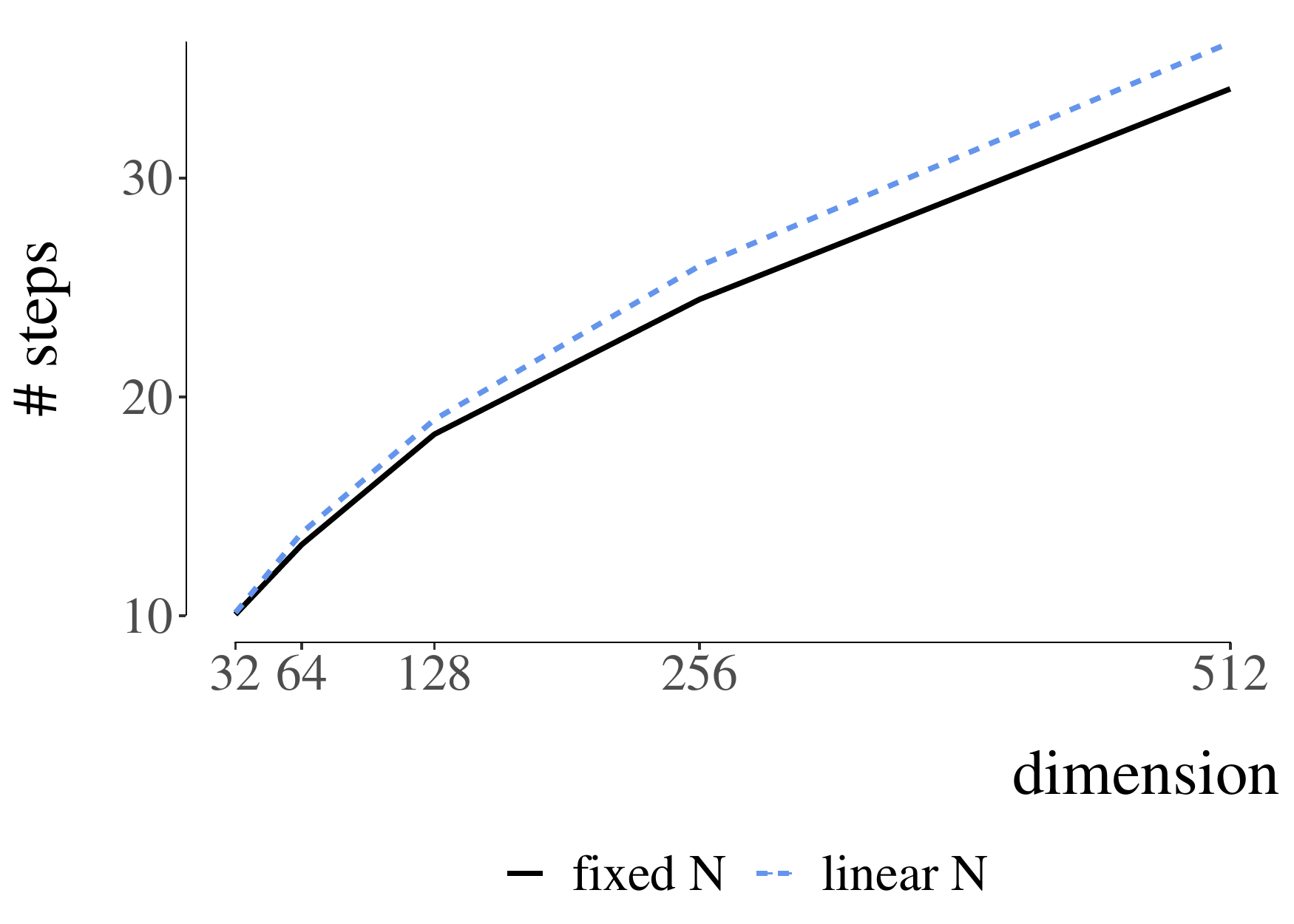}
        \caption{\label{fig:mvnorm:nsteps} Number of bridging distributions.}
    \end{subfigure}
    ~ 
    \begin{subfigure}[t]{0.45\textwidth}
        \centering
        \includegraphics[width=\textwidth]{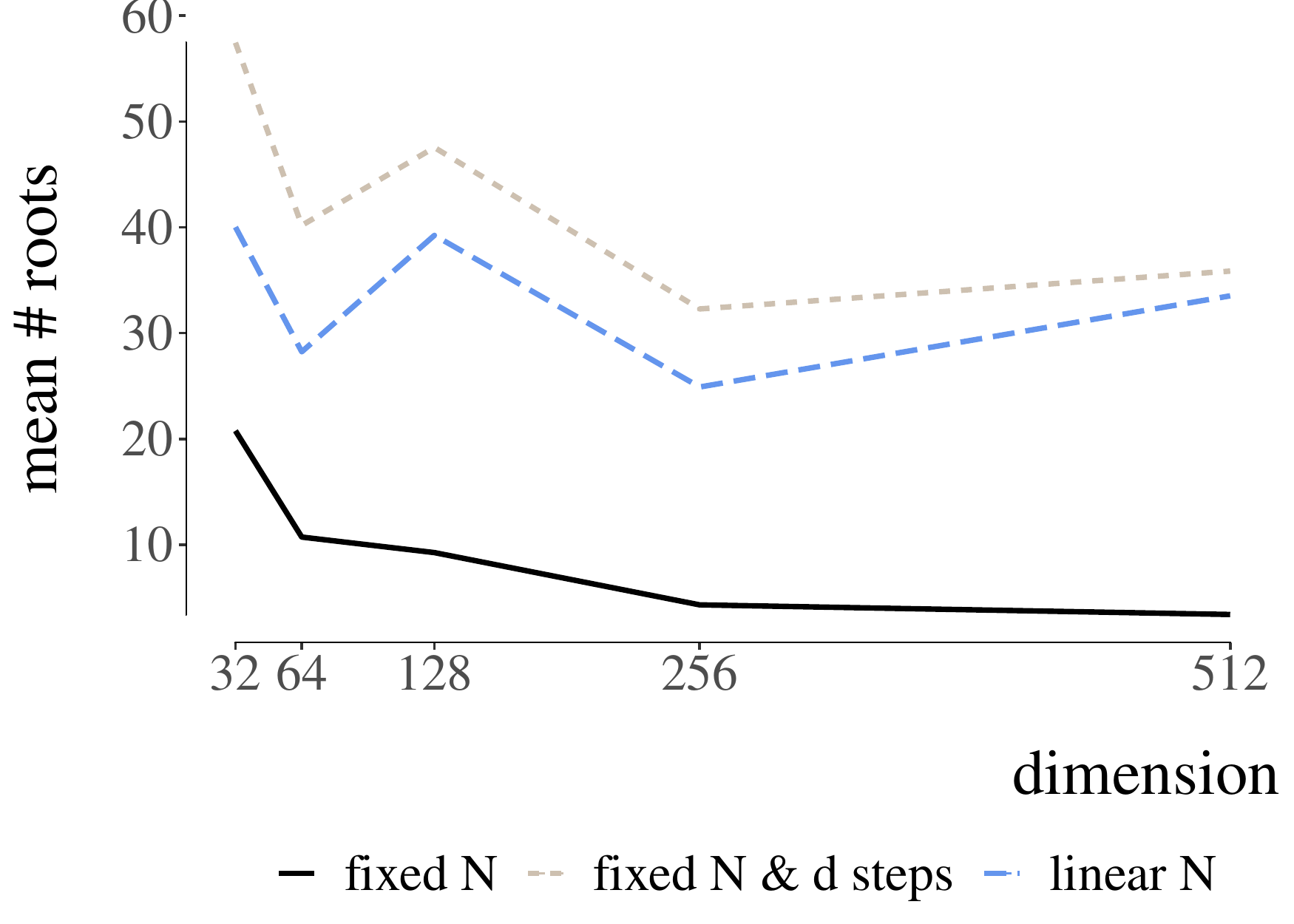}
        \caption{\label{fig:mvnorm:nroots} Number of roots.}
    \end{subfigure}
    \caption{\label{fig:mvnorm:scalings} 
    Normal example of Section \ref{subsec:experiments}. 
    Number of distributions $T$ chosen by an adaptive SMCS 
     (\emph{left}). 
    Number of roots in the genealogical trees, in different regimes (\emph{right}). 
    Plots are obtained by averaging $100$ independent repeats. 
  }
\end{figure*}

Figure \ref{fig:mvnorm:nsteps} shows the number of bridging distributions $T$
obtained using the adaptive strategy in Section \ref{subsec:choosenexttemp},
with an ESS threshold of $\kappa=0.5$. 
The two lines correspond to having $N=256$ (``fixed N'') and $N=256 + 8 d$ (``linear N'') number of particles.
The resulting $T$ appears to increase sub-linearly with $d$ in both regimes. 
We introduce a setup referred to as ``fixed N \& d steps''
in the plots, where $N=256$ and $T = d$; thus inserting more intermediate
distributions than required for controlling the ESS. 
In this setup, $(\lambda_t)_{t\in[T]}$ was determined by 
interpolating between the inverse temperatures obtained from an adaptive SMCS run.
The interpolation was performed using the \texttt{cobs} package in R
\citep{ng2007fast}, which allows one to fit splines constrained to be monotonically 
increasing. 
In the ``fixed N \& d steps'' setup, 
once the sequence $(\lambda_t)_{t\in[T]}$ is obtained,
we run SMCS with adaptive tuning of the MCMC kernels,
we store all quantities required to define these kernels
and we re-run SMCS with these kernels fixed.
That last run generates unbiased estimators of $\hat{Z}$.
Implementation details can be followed in the accompanying R scripts.

To compare the three setups
and also illustrate some of the discussion in Section \ref{sec:theory}
of the article,
we consider the number of ``roots'' or unique ancestors in the genealogical trees 
of the particle systems and plot its average over repeated runs against dimension in Figure \ref{fig:mvnorm:nroots}. 
We observe that the number of roots at the terminal time 
decreases in the ``fixed N'' regime. 
However, it appears stable when either $N$ increases linearly with $d$
in the adaptive SMCS, or when $T$ scales linearly with $d$ for fixed $N$. 
This suggests that the sampler is stable with $d$ 
in these two regimes, where either the number of particles or the number of intermediate steps increases adequately. 

\begin{figure*}[t!]
    \centering
    \begin{subfigure}[t]{0.45\textwidth}
        \centering
        \includegraphics[width=\textwidth]{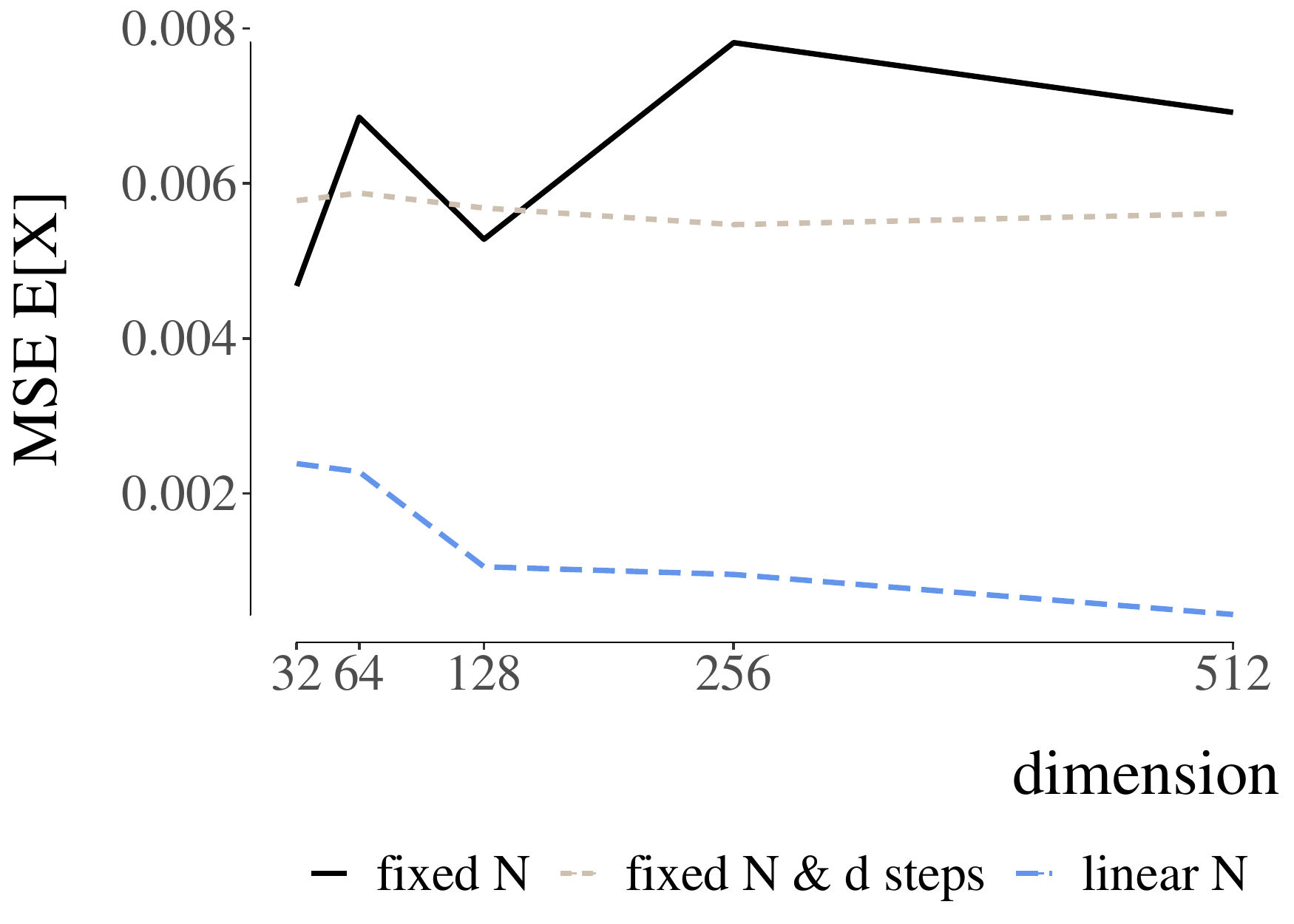}
        \caption{MSE for estimator of $\mathbb{E}_{\pi}[X]$. \label{fig:mvnorm:mse}}
    \end{subfigure}
    ~ 
    \begin{subfigure}[t]{0.45\textwidth}
        \centering
        \includegraphics[width=\textwidth]{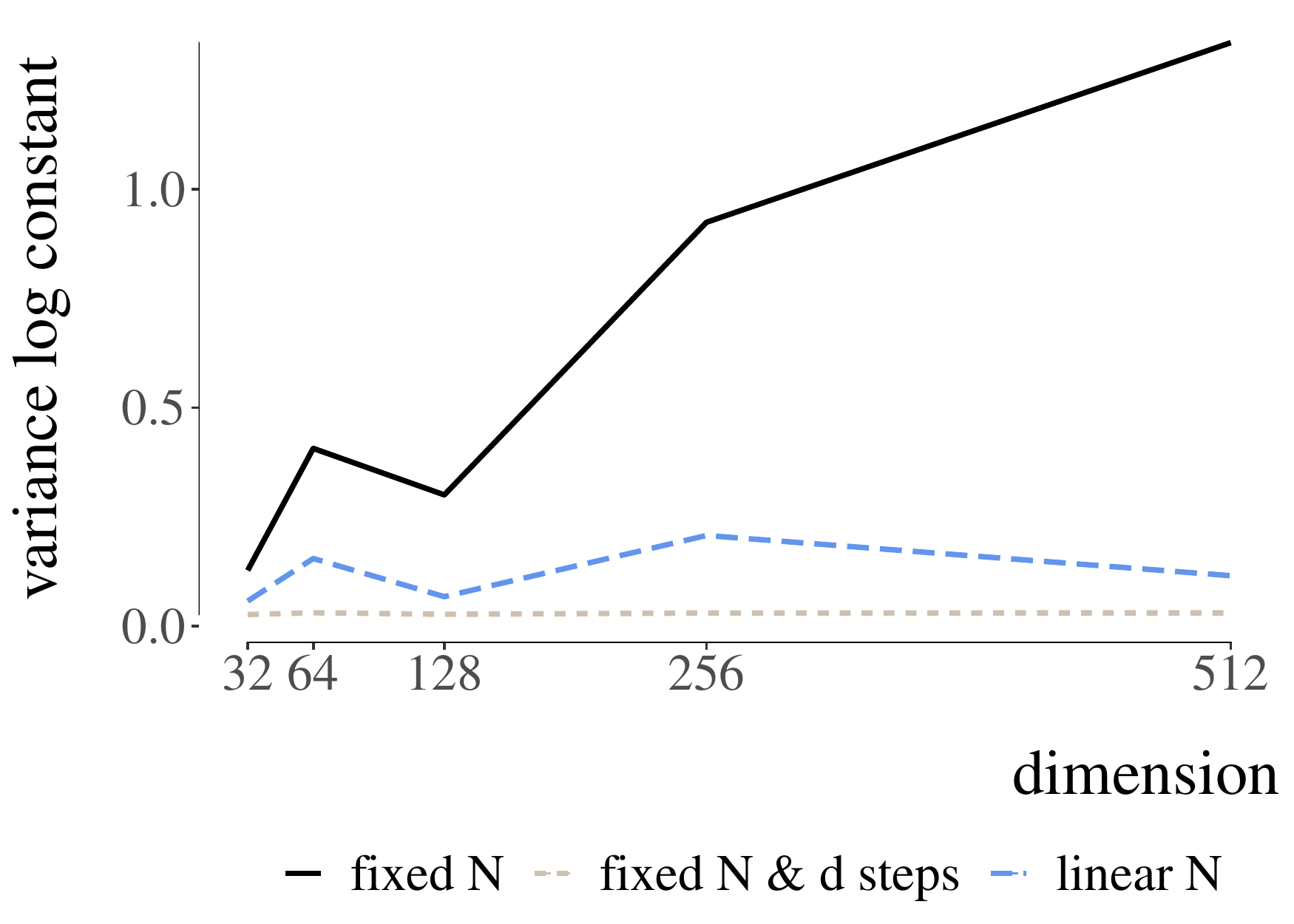}
        \caption{Variance of $\log Z_T^N$. \label{fig:mvnorm:varzhat}}
    \end{subfigure}
    \caption{\label{fig:mvnorm:error} 
    Normal example of Section \ref{subsec:experiments}. 
    MSE for estimator of $\mathbb{E}_{\pi}[X]$ (\emph{left}) and variance of 
    $\log {Z}_T^N$ (\emph{right}) with increasing dimension. 
    Plots are obtained by averaging $100$ independent repeats. }
\end{figure*}

We investigate these regimes further using common measures of Monte Carlo error 
in Figure \ref{fig:mvnorm:error},
such as the 
mean squared error (MSE) associated with the SMCS estimator of $\mathbb{E}_{\pi}[X]=\int_{\mathsf{X}}x\pi(dx)$ 
(averaged over all components in Figure \ref{fig:mvnorm:mse}).
The MSE appears stable in all regimes and even decreases
when $N$ increases with $d$. 
Figure \ref{fig:mvnorm:varzhat} displays the variance of the logarithm of the normalizing
constant estimator $Z^N_T$ against dimension, 
which appears to increase with $d$ for ``fixed N'', but looks stable in the other two
regimes, matching the behaviour of the number of roots in Figure \ref{fig:mvnorm:nroots}.  
These results confirm that SMCS can
indeed deliver a stable accuracy for a polynomial cost 
as the dimension $d$ increases, in settings where importance sampling would fail.

\section{Details concerning the experiments of the article} \label{supplement:detailsexperiments}

We provide the implementation details of SMCS that was used to generate the figures in Section \ref{sec:objects} of the article.
The problem settings are summarized in Table \ref{table:dataset_model}. 

To obtain Figures \ref{fig:merging:beta1}-\ref{fig:merging:beta8} we proceeded as follows.
We assimilate the data in batches of size $10$, and we introduce
intermediate distributions using a geometric path between successive partial posteriors
$p(d\beta|x_{1:10t},y_{1:10t})$ and 
$p(d\beta|x_{1:10(t+1)},y_{1:10(t+1)})$, 
with the convention that $p(d\beta|x_{1:10t},y_{1:10t})$ is $p(d\beta)$
(the prior distribution) for $t=0$.
Each SMCS run employs $N=1024$ particles,
an ESS threshold of $\kappa=0.5$ to determine intermediate distributions
between two partial posteriors, and $2$ HMC iterations per move step
with $\varepsilon = 0.3\times d^{-1/4}$ and $\lceil \varepsilon^{-1} \rceil$ leap-frog steps.
Backward kernels are taken as time reversals of the forward kernels, and 
multinomial resampling was employed.

To obtain Figures \ref{fig:pred1}-\ref{fig:pred2}
we proceeded similarly but with batches of data of size 100. Tuning choices was otherwise identical to the above description.

Moving on to Section \ref{subsec:SIR}, 
we employed Stan \citep{carpenter2017stan} to obtain evaluations of the target density 
and its gradient. The function \texttt{integrate\_ode\_rk45} of Stan was used to solve the SIR ordinary 
differential equation. 
To obtain the partial posteriors shown in Figures \ref{fig:boardingschool:phiinv}-\ref{fig:boardingschool:contours}
we first ran a basic MCMC algorithm targeting the posterior distribution given the first three observations.
Specifically we employed a Metropolis--Rosenbluth--Teller--Hastings algorithm 
with Normal random walk proposals, using a covariance matrix adapted during initial runs,
and using a diagonal matrix with entries $0.01$ in the very first run, along with the starting state 
$\log \gamma = -1$, $\log \beta = 1$, $\log \phi_{\text{inv}}=-1$.
We then calibrated a $3$-dimensional Normal distribution using the mean and variance of the MCMC samples,
to define $\pi_0$. Specifically we obtained
\[\pi_0(\log \gamma, \log \beta, \log \phi_{\text{inv}}) = \mathcal{N}\left(\begin{pmatrix}
  -0.97 \\
  0.49\\
  -2.6
\end{pmatrix},
\begin{pmatrix}
  1.59& 0.25& 0.16 \\ 
  0.25& 0.06& 0.04\\
  0.16& 0.04& 1.08
\end{pmatrix}\right),\]
for the transformed parameters $(\log \gamma, \log \beta, \log \phi_{\text{inv}})$. 
We employed geometric paths interpolating between $\pi_0$ and the posterior distribution given
three observations, and then between successive partial posteriors,
assimilating observations one by one. We ran SMCS
with $N=512$ particles, ESS threshold of $\kappa = 0.5$, $2$ 
HMC iterations per move step, with stepsize of $\varepsilon = 0.1$ and $10$ leap-frog steps. 
Each backward kernel was set as the time reversal of the associated forward kernel.
Multinomial resampling was employed.

To obtain the marginal likelihood plot in Figure \ref{fig:boardingschool:logz},
we employed SMCS using the path of partial posteriors,
with geometric paths interpolating between successive partial posteriors, 
and started from $\pi_0$, exactly as described above. The only difference
is that we used $N=4096$ particles and 3 HMC iterations per move step, resulting in a smaller
variation across 5 independent runs.

\begin{table}
{\scriptsize
\begin{tabular}{ccccc}
\hline 
\hline 
 &  & \textbf{Section 5.1} &  & \textbf{Section 5.2}\tabularnewline
\cline{3-3} \cline{5-5} 
\multirow{3}{*}{Data set} &  & Forest cover type data set &  & English boarding school data set\tabularnewline
 &  & with covariates $x=(x_{1},\ldots,x_{m})$ &  & with daily counts $y=(y_{1},\ldots,y_{m})$\tabularnewline
 &  & and cover type $y=(y_{1},\ldots,y_{m})$ &  & of pupils confined to bed\tabularnewline
 &  &  &  & \tabularnewline
\multirow{2}{*}{Model} &  & Logistic regression model &  & Deterministic SIR model $(S_{t},I_{t},R_{t})_{t\geq t_{0}}$\tabularnewline
 &  & $p(y|x,\beta)=\prod_{i=1}^{m}\mathcal{B}(y_{i};(1+\exp(-x_{i}^{\top}\beta))^{-1})$ &  & $p(y|t_{0},\theta)=\prod_{t=1}^{m}\mathcal{NB}(y_{t};I_{t},\phi_{\text{inv}})$\tabularnewline
 &  &  &  & \tabularnewline
Parameters &  & $\beta$ contains regression coefficients &  & $\theta=(\gamma,\beta,\phi_{\text{inv}})$ contains infection rate,\tabularnewline
 &  &  &  & recovery rate and dispersion\tabularnewline
 &  &  &  & \tabularnewline
Prior distribution &  & $p(\beta)=\prod_{i=1}^d\mathcal{N}(\beta_i;0,10)$ &  & $p(\theta)=\mathcal{TN}(\gamma;0.4,0.5^2)\mathcal{TN}(\beta;2,1)\mathcal{E}(\phi_{\text{inv}};5)$\tabularnewline
 &  &  &  & \tabularnewline
Target distribution $\pi$ &  & Posterior distribution $\pi(\beta)=p(\beta|x,y)$ &  & Posterior distribution $\pi(\theta)=p(\theta|t_{0},y)$\tabularnewline
 &  &  &  & \tabularnewline
Normalizing constant $Z$ &  & Marginal likelihood $Z=p(y|x)$ &  & Marginal likelihood $Z(t_{0})=p(y|t_{0})$\tabularnewline
\hline 
\hline 
\end{tabular}
}
\caption{Summary of examples in Section \ref{sec:objects} of the article. $\mathcal{B}$ denotes the Bernoulli distribution, $\mathcal{NB}$ the Negative Binomial distribution, $\mathcal{TN}$ the Truncated Normal distribution with support on $\mathbb{R}_+$ and $\mathcal{E}$ the exponential distribution. \label{table:dataset_model}}
\end{table}

\section{Logistic regression: different paths}

\begin{figure*}[t!]
    \centering
    \begin{subfigure}[t]{0.3\textwidth}
        \centering
        \includegraphics[width=0.8\textwidth]{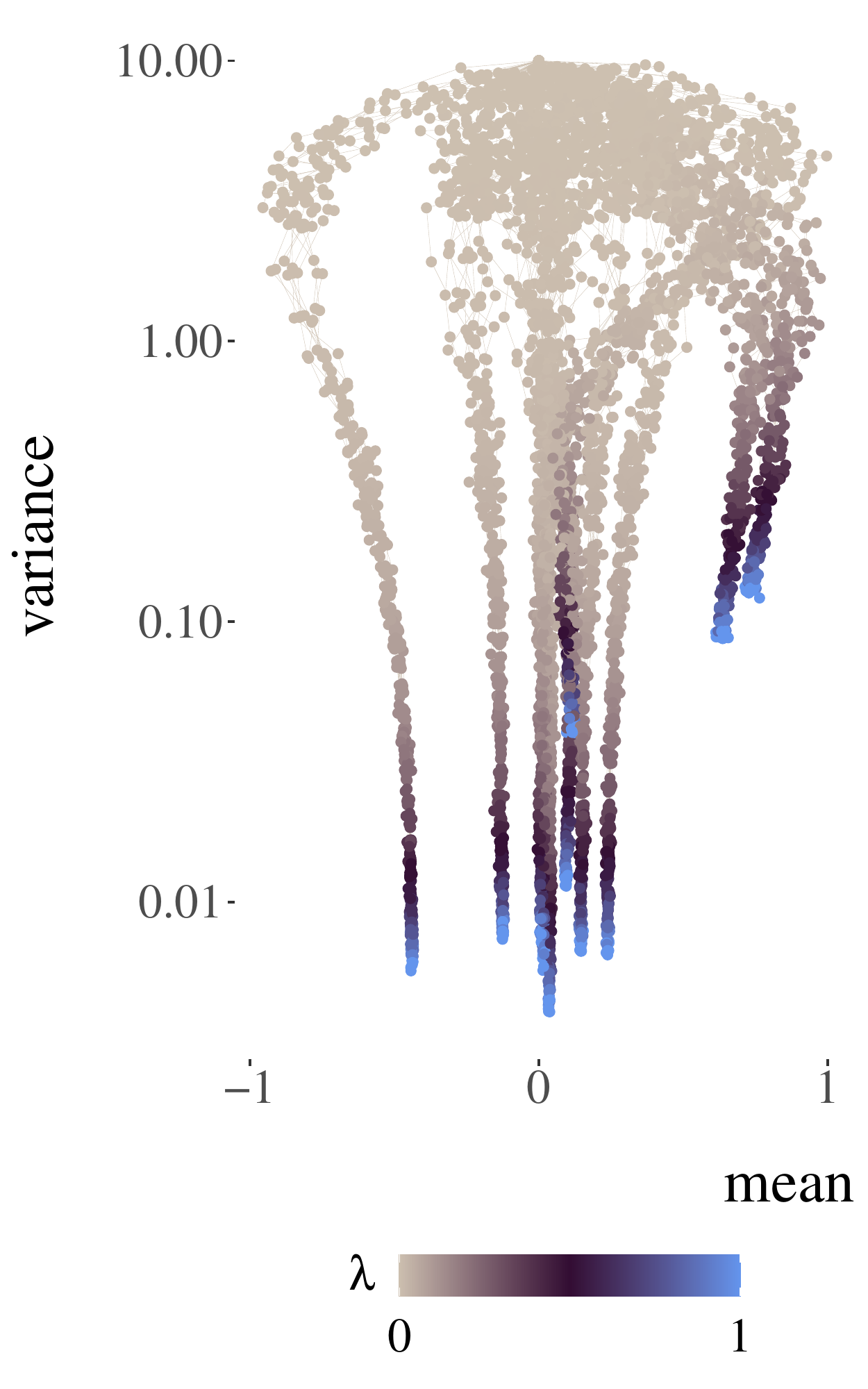}
        \caption{Geometric path.\label{fig:path:geometric}}
    \end{subfigure}
    ~ 
    \begin{subfigure}[t]{0.3\textwidth}
        \centering
        \includegraphics[width=0.8\textwidth]{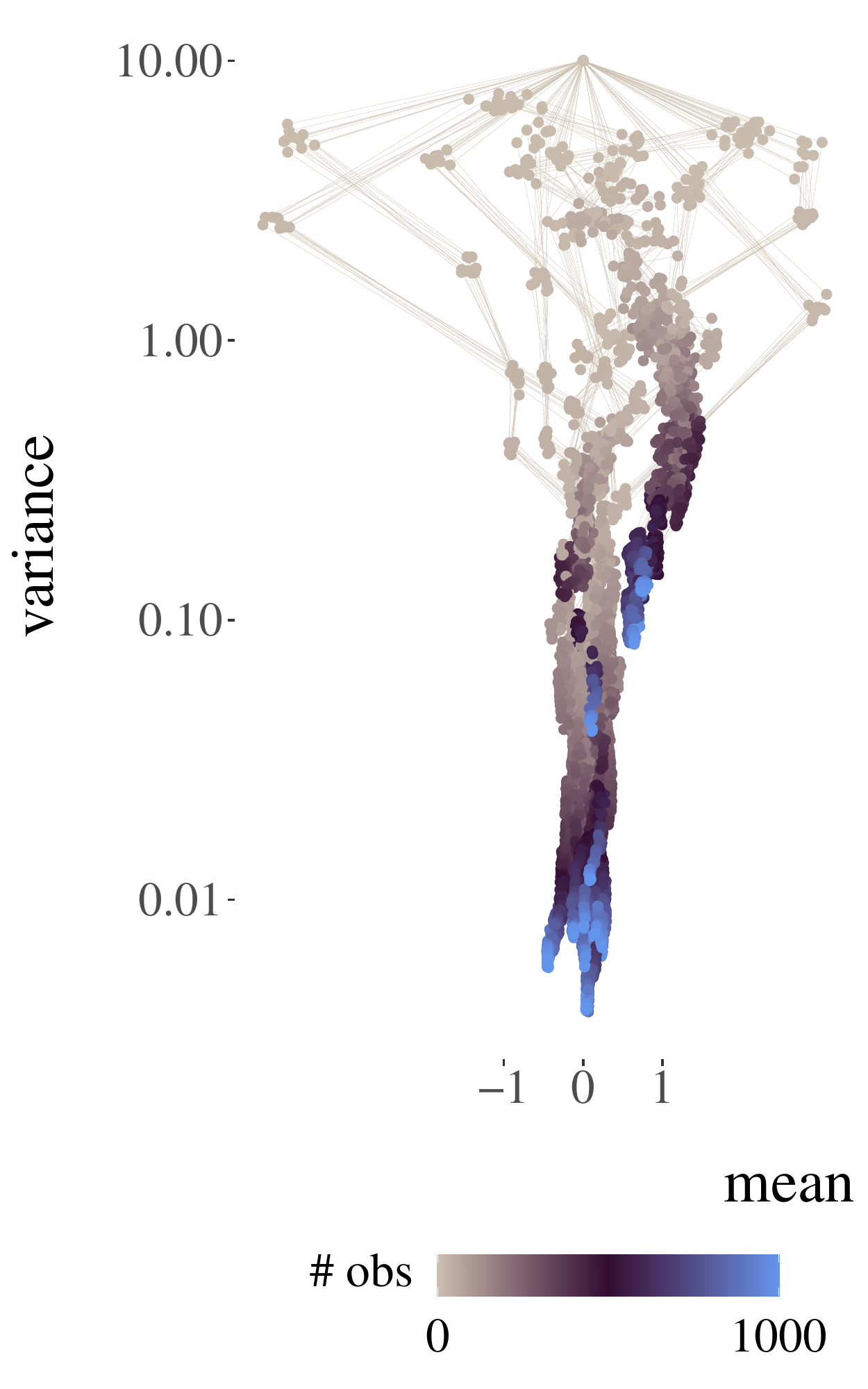}
        \caption{Partial posteriors.\label{fig:path:partial}}
    \end{subfigure}
    ~
    \begin{subfigure}[t]{0.3\textwidth}
        \centering
        \includegraphics[width=0.8\textwidth]{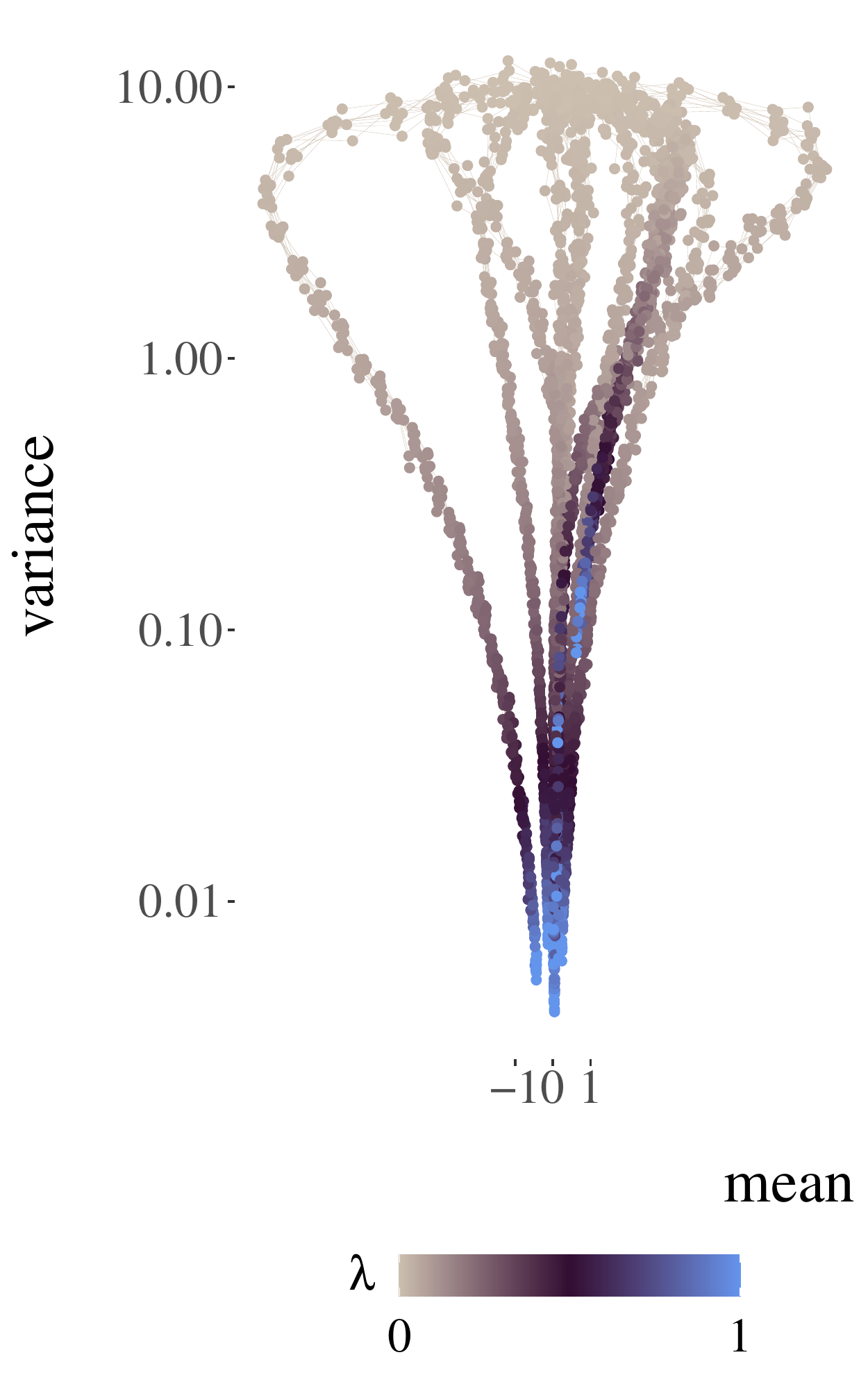}
        \caption{Path of least coding.\label{fig:path:pgg}}
    \end{subfigure}
    \caption{\label{fig:path} Three paths of distributions connecting the prior
    to the posterior in a logistic regression example described in Section \ref{subsec:logistic} of the article,
  represented by lines in the mean-variance plane, for each component
  and 10 independent runs.}
\end{figure*}

We consider the logistic regression example described in Section \ref{subsec:logistic} of the article,
and mentioned also in Section \ref{subsec:choosepath}.
We used $N=1024$ particles and an ESS threshold of $\kappa=0.5$.
Using the first $m=1000$ rows of the data,
Figure \ref{fig:path} shows the mean and variance of the $d=11$ components of $\beta$
for three paths of distributions: a geometric path (\ref{fig:path:geometric}),
a path of partial posteriors where observations are assimilated in batches of size 10 (\ref{fig:path:partial}),
and a path of ``least coding effort''  using the P{\'o}lya--Gamma Gibbs sampler (\ref{fig:path:pgg})
with scaled covariates $\lambda_t x$ and $\lambda_t \in[0,1]$. 
HMC moves were employed for the first two paths, with stepsize $\varepsilon = 0.3\times d^{-1/4}$, $\lceil \varepsilon^{-1} \rceil$ leap-frog steps, and 10 independent repeats are shown.
In all cases, we set the backward kernels to be the time reversal of the corresponding forward kernels, and employed multinomial resampling for all simulations.
The three paths start and end at the same distributions 
but the intermediate distributions are visibly different.

\section{Partial posterior averaged over orderings}

We illustrate the usefulness of unbiased estimators
in the setting of logistic regression,
as in Section \ref{subsec:logistic}
of the article. 
As described in \citet{middleton2019unbiased}, unbiased estimators can be generically
obtained from SMCS through the coupling of an MCMC scheme 
proposed in \citet{andrieu2010particle}, known as ``particle independent Metropolis--Hastings'' (PIMH).

Suppose that one of the covariates in the regression
is actually a random draw from another model, as in two-step estimation
\citep{murphy2002estimation}, and that we want to propagate that uncertainty 
onto the posterior. 
In the Bayesian terminology, this could lead to a
``cut distribution'' \citep{plummer2015cuts}. The same computational challenge occurs when
some data are missing and multiple imputation is performed \citep{rubin1996multiple},
or in the context of propensity scores \citep{zigler2014uncertainty}, 
or when bagging posteriors \citep{buhlmann2014discussion}.
All these cases are instances of the generic question of approximating $\pi(dx) = \int \pi(dx|\eta) g(d\eta)$, 
where we can sample from $\eta\sim g$ and design a Monte Carlo method to 
approximate the conditional distribution $\pi(dx|\eta)$. 
With unbiased approximations of $\pi(dx|\eta)$ for any $\eta \sim g$,
we obtain an unbiased approximation of $\pi(dx)$ itself. 

\begin{figure*}[t!]
    \centering
        \includegraphics[width=0.5\textwidth]{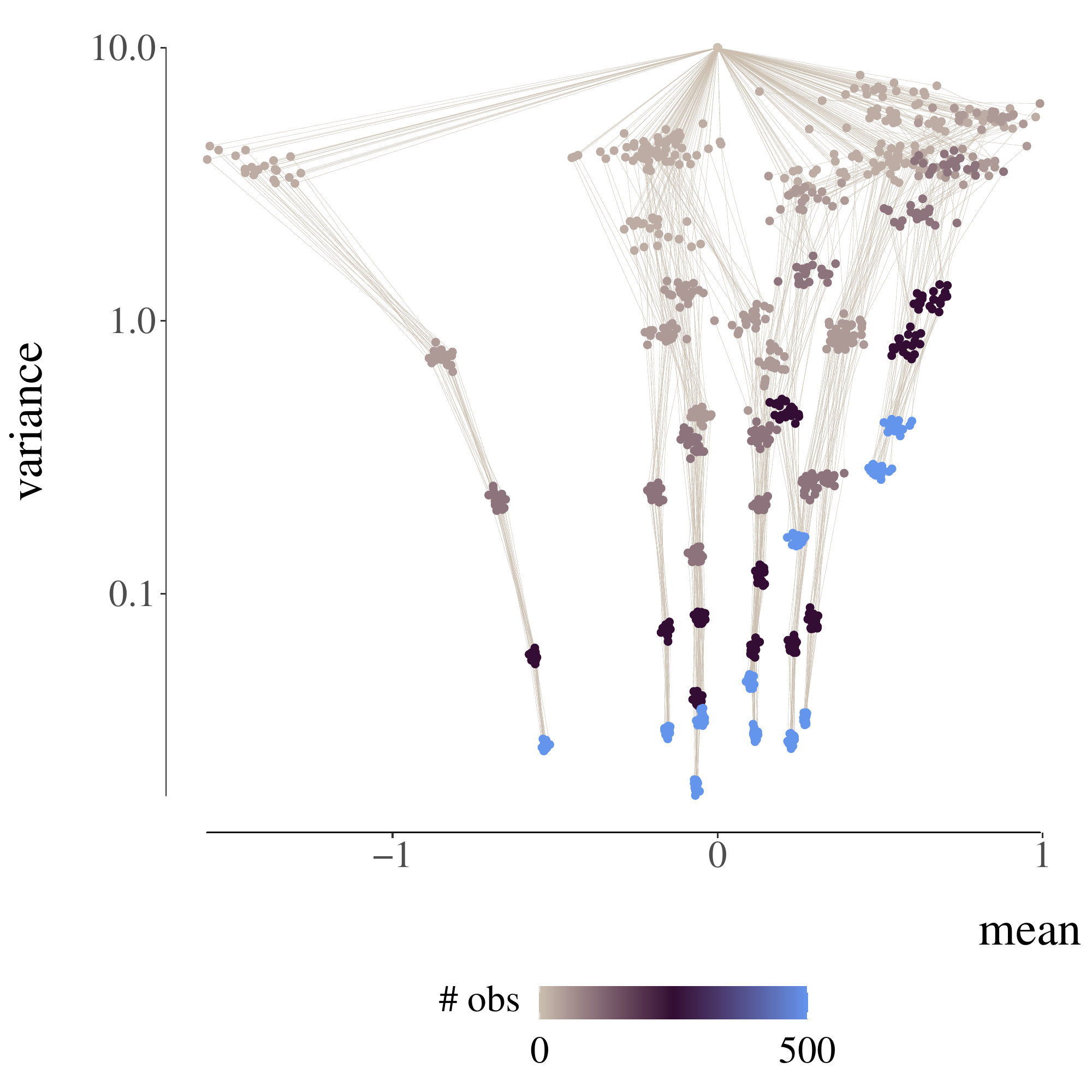}
    \caption{\label{fig:path:partialaverage} Logistic regression with forest cover type data. 
    Path of partial posteriors 
    averaged over orderings of the data. The ten realizations
are obtained by non-parametric bootstrap from $R=1000$ independent unbiased estimators. 
}
\end{figure*}

Here we consider a variant of the
path of partial posteriors for logistic regression, $\pi_m(d\beta)=p(d\beta|x_{1:m},y_{1:m})$ given $m$
observations. This path usually depends
on a specific ordering of the observations. As an alternative 
we consider the
posterior distribution averaged over orderings, 
$\pi^\star_m(d\beta) = (m!)^{-1}\sum_{\sigma} p(d\beta|
x_{\sigma(1:m)},y_{\sigma(1:m)})$ where $\sigma(1:m)$ denotes the first $m$ 
elements of a permutation of the entire data set.
To obtain
unbiased estimators, we sample $m$ observations from the data set at random
without replacement, and run coupled PIMH chains 
\citep{middleton2019unbiased} targeting $p(d\beta|
x_{\sigma(1:m)},y_{\sigma(1:m)})$ using SMCS with $N=128$ particles 
as proposals. Each SMCS run employs HMC moves with stepsize $\varepsilon = 0.3\times d^{-1/4}$
and $3$ leap-frog steps. The schedule and the mass matrices 
are obtained using an initial run of adaptive SMCS on a data set of the same size $m$, and 
these tuning parameters are then fixed
in the generation of unbiased estimators. 
Indeed the use of adaptive techniques would jeopardize the lack-of-bias property
of $Z^N_T$ and thus could change the target distribution of the PIMH kernels.
We compute $R=1000$ independent unbiased estimators for each $m$, 
and use empirical averages to approximate $\pi^\star_m(d\beta)$. 
Figure \ref{fig:path:partialaverage} illustrates the evolution of the means and 
variances of $\beta$ as $m$ increases. These distributions can be interpreted
as the posterior distributions given $m$ randomly chosen observations from the data set,
as opposed to conditioning on the first $m$ observations in an arbitrary order.

\section{Laplace approximations to initialize SMCS \label{supplement:laplace}}

\begin{figure*}[ht!]
    \centering
    \begin{subfigure}[t]{0.45\textwidth}
        \centering
        \includegraphics[width=\textwidth]{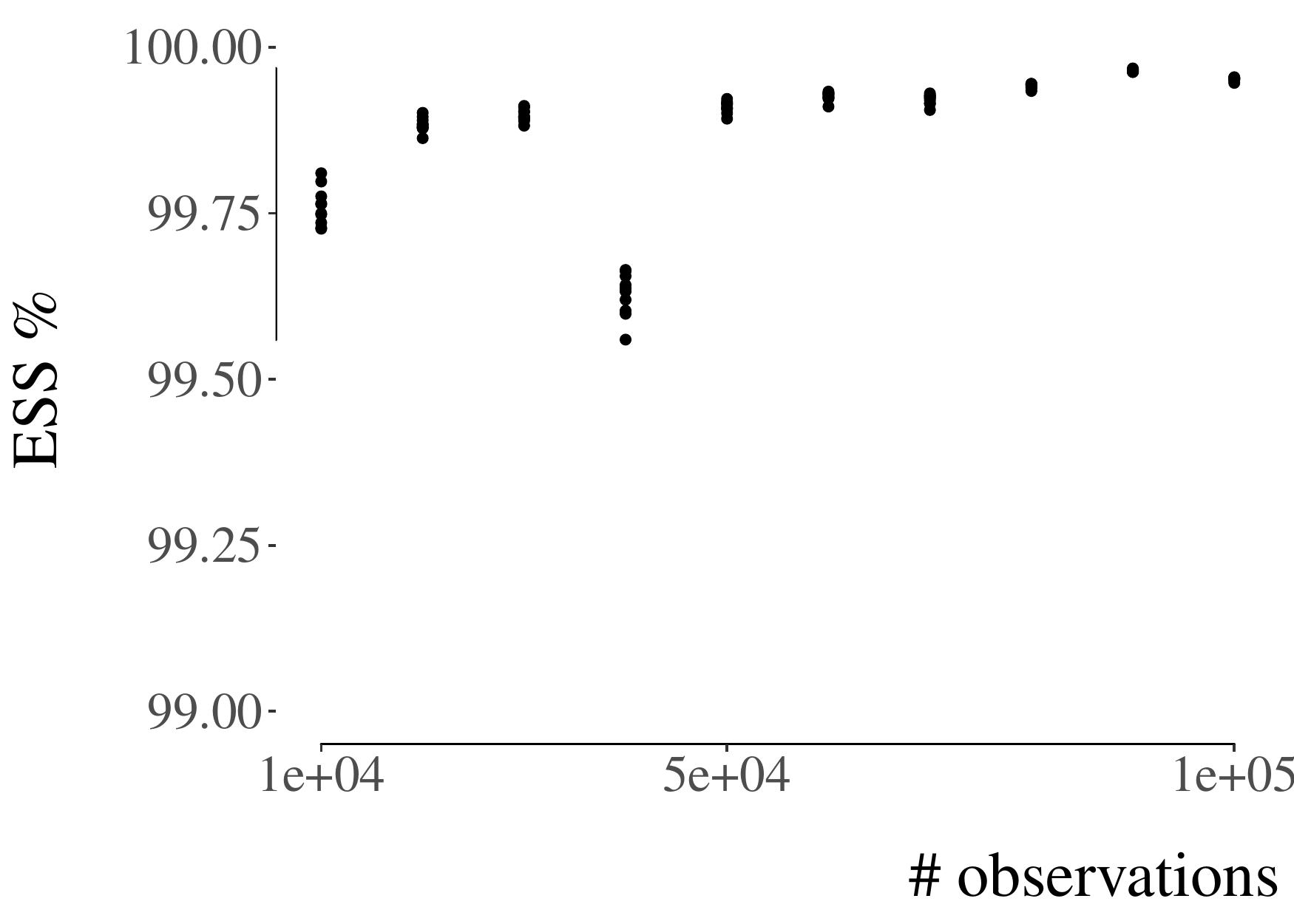}
        \caption{\label{fig:logistic:ess} Effective sample size.}
    \end{subfigure}
    ~
    \begin{subfigure}[t]{0.45\textwidth}
        \centering
        \includegraphics[width=\textwidth]{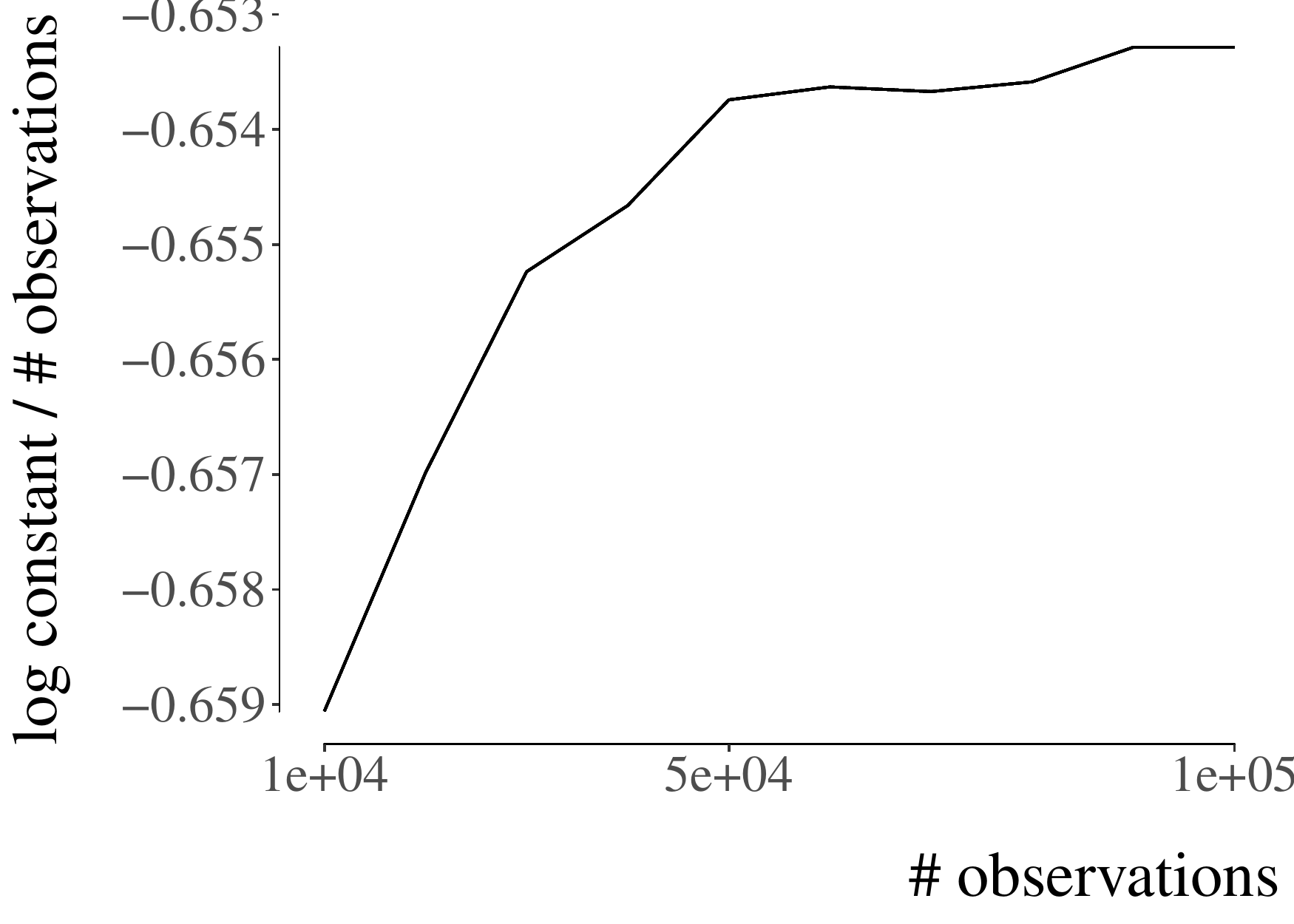}
        \caption{\label{fig:logistic:estimates} Estimates of $\log Z/m$.}
    \end{subfigure}
    \caption{\label{fig:laplace} Logistic regression with forest cover type data. 
    ESS against number of observations $m$ (\emph{left}) and 
    estimates of $\log Z/m$ (\emph{right}), when initializing from a Laplace approximation of the posterior.}
\end{figure*}

Bayesian asymptotics provide useful strategies for Monte Carlo
computation.  For example, we can use a Laplace
approximation of the posterior as initial distribution $\pi_0$, i.e. a Normal 
distribution centered at the maximum likelihood estimate (MLE) and with covariance given by the inverse of the 
information matrix at the MLE \citep[e.g.][]{chopin2017leave}. 

In the context of Section \ref{subsec:logistic}, 
Figure \ref{fig:logistic:ess} shows that the
approximation is extremely accurate when the number of observations $m$ is large and leads to very
high effective sample sizes in a single step of importance sampling.
Here SMCS employs $N=1024$ particles, an ESS threshold of $\kappa=0.5$ and HMC moves,
with step size $\varepsilon = 0.3\times d^{-1/4}$ and $\lceil \varepsilon^{-1}\rceil$ leap-frog steps.
The relative variance of $Z_T^N$ is close to zero for data sizes above $m=10^4$.
Thus SMCS that employ the ESS 
criterion to select the next inverse temperature would default back to plain importance sampling.
Figure \ref{fig:logistic:estimates} shows the estimates of $\log Z$ 
divided by the number of observations $m$; ten repeats are overlaid but the accuracy is such that
 they are indistinguishable. 
Some effects of Bayesian asymptotics on the performance of 
SMCS are studied in \citet{chopin2002sequential},
samplers for tall data settings are proposed in \citet{gunawan2018subsampling},
and approximate samplers for sequential inference in \citet{gerber2020online}.

\end{document}